\documentclass[11pt]{article}
\usepackage{amsmath,amssymb}

\def\baselinestretch{1.2}
\catcode`\@=11
\def\@citex[#1]#2{%
\if@filesw \immediate \write \@auxout {\string \citation {#2}}\fi
\@tempcntb\m@ne \let\@h@ld\relax \def\@citea{}%
\@cite{%
  \@for \@citeb:=#2\do {%
    \@ifundefined {b@\@citeb}%
      {\@h@ld\@citea\@tempcntb\m@ne{\bf ?}%
      \@warning {Citation `\@citeb ' on page \thepage \space undefined}}%
      {\@tempcnta\@tempcntb \advance\@tempcnta\@ne%
      \@tempcntb\number\csname b@\@citeb \endcsname \relax%
      \ifnum\@tempcnta=\@tempcntb %Number follows previous--hold on to it
        \ifx\@h@ld\relax%
          \edef \@h@ld{\@citea\csname b@\@citeb\endcsname}%
        \else%
          \edef\@h@ld{\ifmmode{-}\else--\fi\csname b@\@citeb\endcsname}%
        \fi%
      \else%   %  non-successor--dump what's held and do this one
        \@h@ld\@citea\csname b@\@citeb \endcsname%
        \let\@h@ld\relax%
      \fi}%
    \def\@citea{,\penalty\@highpenalty\,}%
  }\@h@ld
}{#1}}

\def\@citeb#1#2{{[#1]\if@tempswa , #2\fi}}
\def\@citeu#1#2{{$^{#1}$\if@tempswa , #2\fi }}
\def\@citep#1#2{{#1\if@tempswa , #2\fi}}

\def\bcites{         % cite with []'s
        \catcode`\@=11
        \let\@cite=\@citeb
        \catcode`\@=12
}

\def\upcites{         % cite with exponents
        \catcode`\@=11
        \let\@cite=\@citeu
        \catcode`\@=12
}

\def\plaincites{      % cite without brackets
        \catcode`\@=11
        \let\@cite=\@citep
        \catcode`\@=12
}

\newcount\hour
\newcount\minute
\newtoks\amorpm
\hour=\time\divide\hour by 60 \minute=\time{\multiply\hour by 60
\global\advance\minute by-\hour}
\edef\standardtime{{\ifnum\hour<12 \global\amorpm={am}%
        \else\global\amorpm={pm}\advance\hour by-12 \fi
        \ifnum\hour=0 \hour=12 \fi
        \number\hour:\ifnum\minute<10 0\fi\number\minute\the\amorpm}}
\edef\militarytime{\number\hour:\ifnum\minute<10 0\fi\number\minute}

\def\draftlabel#1{{\@bsphack\if@filesw {\let\thepage\relax
   \xdef\@gtempa{\write\@auxout{\string
      \newlabel{#1}{{\@currentlabel}{\thepage}}}}}\@gtempa
   \if@nobreak \ifvmode\nobreak\fi\fi\fi\@esphack}
        \gdef\@eqnlabel{#1}}
\def\@eqnlabel{}
\def\@vacuum{}
\def\marginnote#1{}
\def\draftmarginnote#1{\marginpar{\raggedright\scriptsize\tt#1}}
\def\draft{
        \pagestyle{plain}
        \overfullrule=2pt
        \oddsidemargin -.5truein
        \def\@oddhead{\sl \phantom{\today\quad\militarytime} \hfil
        \smash{\Large\sl DRAFT} \hfil \today\quad\militarytime}
        \let\@evenhead\@oddhead
        \let\label=\draftlabel
        \let\marginnote=\draftmarginnote
        \def\ps@empty{\let\@mkboth\@gobbletwo
        \def\@oddfoot{\hfil \smash{\Large\sl DRAFT} \hfil}
        \let\@evenfoot\@oddhead}
        \def\@eqnnum{(\theequation)\rlap{\kern\marginparsep\tt\@eqnlabel}%
        \global\let\@eqnlabel\@vacuum}  }

\def\blackfonts{
        \font\blackboard=msbm10 scaled\magstep1
        \font\blackboards=msbm8
        \font\blackboardss=msbm6
}

\def\prep{         % twocolumn.sty  Changed by Marek and Neil
        \catcode`\@=11
        \input art10.sty
        \catcode`\@=12
        
        \let\small\null
        \def\blackfonts{
                \font\blackboard=msbm10
                \font\blackboards=msbm7
                \font\blackboardss=msbm5
        }
        \let\sl\it
        \twocolumn
        \sloppy
        \voffset=-2.54truecm
        \hoffset=-2.54truecm
        \flushbottom
        \parindent 1em
        \leftmargini 2em
        \leftmarginv .5em
        \leftmarginvi .5em
        \marginparwidth 48pt
        \marginparsep 10pt
        \setlength{\columnsep}{2truecm}
        \setlength{\textwidth}{25.4truecm}
        \setlength{\textheight}{17truecm}
        \baselineskip=16pt
        \oddsidemargin .18truein
        \evensidemargin .17truein
}

\def\eqalign#1{\null\,\vcenter{\openup\jot\m@th
  \ialign{\strut\hfil$\displaystyle{##}$&$\displaystyle{{}##}$\hfil
      \crcr#1\crcr}}\,}
\def\eqalignno#1{\displ@y \tabskip\centering
  \halign to\displaywidth{\hfil$\@lign\displaystyle{##}$\tabskip\z@skip
    &$\@lign\displaystyle{{}##}$\hfil\tabskip\centering
    &\llap{$\@lign##$}\tabskip\z@skip\crcr
    #1\crcr}}

\def\section{\@startsection {section}{1}{\z@}{3.ex plus 1ex minus
 .2ex}{2.ex plus .2ex}{\large\bf}}
\def\subsection{\@startsection{subsection}{2}{\z@}{2.75ex plus 1ex minus
 .2ex}{1.5ex plus .2ex}{\bf}}

\def\appendix{{\newpage\section*{Appendix}}\let\appendix\section%
        {\setcounter{section}{0}
        \gdef\thesection{\Alph{section}}}\section}

\def\abstract{\if@twocolumn
\section*{Abstract}
\else %\small
\begin{center}
{\bf Abstract\vspace{-.5em}\vspace{0pt}}
\end{center}
\quotation \fi}

\catcode`\@=12
\def\d{\partial}

\def\sqr#1#2{{\vcenter{\vbox{\hrule height.#2pt\hbox{\vrule width.#2pt
height#1pt \kern#1pt \vrule width.#2pt}\hrule height.#2pt}}}}
\def\w{\mathchoice\sqr45\sqr45\sqr{2.1}3\sqr{1.5}3\,}

\def\=d{\,{\buildrel\rm def\over =}\,}

\def\F{{\cal F}}

\def\i3p{\p32\int d^3p}

\def\As{A\hbox to 1pt{\hss /}}
\def\np4{\int d^4p_1\cdots d^4p_{n-1}\, }

\def\Tr{{\rm Tr}\, }

\def\nx4{\int d^4x_1\ldots d^4x_n\, }

\def\kon#1#2{\vbox{\halign{##&&##\cr
\lower4pt\hbox{$\scriptscriptstyle\vert$}\hrulefill &
\hrulefill\lower4pt\hbox{$\scriptscriptstyle\vert$}\cr $#1$& $#2$\cr}}}

\def\konv#1#2#3{\hbox{\vrule height12pt depth-1pt}
\vbox{\hrule height12pt width#1cm depth-11.6pt} \hbox{\vrule height6.5pt
depth-0.5pt} \vbox{\hrule height11pt width#2cm depth-10.6pt\kern5pt
      \hrule height6.5pt width#2cm depth-6.1pt}
\hbox{\vrule height12pt depth-1pt} \vbox{\hrule height6.5pt width#3cm
depth-6.1pt} \hbox{\vrule height6.5pt depth-0.5pt}}
\def\konu#1#2#3{\hbox{\vrule height12pt depth-1pt}
\vbox{\hrule height1pt width#1cm depth-0.6pt} \hbox{\vrule height12pt
depth-6.5pt} \vbox{\hrule height6pt width#2cm depth-5.6pt\kern5pt
      \hrule height1pt width#2cm depth-0.6pt}
\hbox{\vrule height12pt depth-6.5pt} \vbox{\hrule height1pt width#3cm
depth-0.6pt} \hbox{\vrule height12pt depth-1pt}}

\def\konw#1#2#3{\hbox{\vrule height12pt depth-1pt}
\vbox{\hrule height12pt width#1cm depth-11.6pt} \hbox{\vrule height6.5pt
depth-0.5pt} \vbox{\hrule height12pt width#2cm depth-11.6pt \kern5pt
      \hrule height6.5pt width#2cm depth-6.1pt}
\hbox{\vrule height6.5pt depth-0.5pt} \vbox{\hrule height12pt width#3cm
depth-11.6pt} \hbox{\vrule height12pt depth-1pt}}

\def\i{{\rm int}}

\def\m3{{\mu_1\mu_2\mu_3}}

\def\p{{(+)}}

\def\be{\begin{equation}}       \def\eq{\begin{equation}}
\def\ee{\end{equation}}         \def\eqe{\end{equation}}

\def\bea{\begin{eqnarray}}      \def\eqa{\begin{eqnarray}}
\def\ena{\end{eqnarray}}        \def\eea{\end{eqnarray}}
                                \def\eqae{\end{eqnarray}}

\def\ba{\begin{array}}
\def\ea{\end{array}}
\def\unit{1 \hskip-.3em \raise2pt\hbox{$ \scriptstyle |$ } }
\def\d{\delta}
\def\g{\gamma}

\def\i{\iota}

\def\l{\lambda}
\def\m{\mu}
\def\n{\nu}
  \def\w{\omega}
\def\p{\pi}                % Also, \varpi
\def\t{\tau}

\def\D{\Delta}
\def\F{\Phi}

\def\bop#1{\setbox0=\hbox{$#1M$}\mkern1.5mu
        \vbox{\hrule height0pt depth.04\ht0
        \hbox{\vrule width.04\ht0 height.9\ht0 \kern.9\ht0
        \vrule width.04\ht0}\hrule height.04\ht0}\mkern1.5mu}
\def\pa{\partial}                              % curly d
\def\>{\rangle} %right angle

\def\<{\langle} %left angle
\def\Dsl{D \hskip-.6em \raise1pt\hbox{$ / $ } }

\def\sl#1{\rlap{\hbox{$\mskip 1 mu /$}}#1}% good slash for l.c.
\def\leftrightarrowfill{$\mathsurround=0pt \mathord\leftarrow \mkern-6mu
       \cleaders\hbox{$\mkern-2mu \mathord- \mkern-2mu$}\hfill
       \mkern-6mu \mathord\rightarrow$}
\def\dvec#1{\vbox{\ialign{##\crcr
       \leftrightarrowfill\crcr\noalign{\kern-1pt\nointerlineskip}
       $\hfil\displaystyle{#1}\hfil$\crcr}}}          % <--> accent
\def\hook#1{{\vrule height#1pt width0.4pt depth0pt}}
\def\leftrighthookfill#1{$\mathsurround=0pt \mathord\hook#1
       \hrulefill\mathord\hook#1$}
\def\underhook#1{\vtop{\ialign{##\crcr                 % |_| under
       $\hfil\displaystyle{#1}\hfil$\crcr
       \noalign{\kern-1pt\nointerlineskip\vskip2pt}
       \leftrighthookfill5\crcr}}}
\def\smallunderhook#1{\vtop{\ialign{##\crcr      % " for su'scripts
       $\hfil\scriptstyle{#1}\hfil$\crcr
       \noalign{\kern-1pt\nointerlineskip\vskip2pt}
       \leftrighthookfill3\crcr}}}
\def\sfrac#1#2{{\vphantom1\smash{\lower.5ex\hbox{\small$#1$}}\over
       \vphantom1\smash{\raise.4ex\hbox{\small$#2$}}}} % alt. fraction
\def\bfrac#1#2{{\vphantom1\smash{\lower.5ex\hbox{$#1$}}\over
       \vphantom1\smash{\raise.3ex\hbox{$#2$}}}}      % "
\def\afrac#1#2{{\vphantom1\smash{\lower.5ex\hbox{$#1$}}\over#2}}  %"
\def\on#1#2{{\buildrel{\mkern2.5mu#1\mkern-2.5mu}\over{#2}}}%acc.over
\def\ddt#1{\on{\hbox{\LARGE .\kern-2pt.}}#1}             % double dot
\def\tdt#1{\on{\hbox{\LARGE .\kern-2pt.\kern-2pt.}}#1}   % triple dot

\def\boxes#1{
       \newcount\num
       \num=1
       \newdimen\downsy
       \downsy=-1.5ex
       \mskip-2.8mu
       \bo
       \loop
       \ifnum\num<#1
       \llap{\raise\num\downsy\hbox{$\bo$}}
       \advance\num by1
       \repeat}
\def\boxup#1#2{\newcount\numup
       \numup=#1
       \advance\numup by-1
       \newdimen\upsy
       \upsy=.75ex
       \mskip2.8mu
       \raise\numup\upsy\hbox{$#2$}}

\newskip\humongous \humongous=0pt plus 1000pt minus 1000pt
\def\caja{\mathsurround=0pt}
\def\eqalign#1{\,\vcenter{\openup2\jot \caja
       \ialign{\strut \hfil$\displaystyle{##}$&$
       \displaystyle{{}##}$\hfil\crcr#1\crcr}}\,}
\newif\ifdtup

\def\to{\rightarrow}

\def\1ov4{{1\over 4}}

\def\Tr{{\rm Tr}}

\def\pa{\partial}

\def\ddt{\dot{\t}}

\def\pa{\partial}

\def\pa{\partial}

\renewcommand{\d}{\delta}

\newcommand{\beq}{\begin{equation}}
\newcommand{\eeq}{\end{equation}}

\def\ba{\begin{eqnarray}}
\def\ea{\end{eqnarray}}

\def\w{\vec{w}}

\begin{document}
%%%%%%%%%
% {}Front page here
\thispagestyle{empty}
%\vspace*{1cm}
\null\vskip-24pt \hfill AEI-2001-138
 \vskip-10pt \hfill  CERN-TH-2001-337
 \vskip-10pt \hfill LAPTH-894/01
 \vskip-10pt \hfill {\tt hep-th/0201145}
\vskip0.2truecm
\begin{center}
\vskip 0.2truecm {\Large\bf
%\titleline
Implications of Superconformal Symmetry\\ for Interacting (2,0) Tensor
Multiplets
%Four-Point Functions and OPE of Stress-Tensor Multiplets\\
%in Superconformal (2,0) Theory
}\\
\vskip 1truecm
%\vfill
{\bf G. Arutyunov$^{*,**}$
 \footnote{email:{\tt agleb@aei-potsdam.mpg.de}},
E. Sokatchev$^{\dagger,\ddagger}$
 \footnote{email:{\tt Emery.Sokatchev@cern.ch} \\
$^{**}$On leave of absence from Steklov Mathematical Institute, Gubkin str.8,
117966, Moscow, Russia   \\
$^{\ddagger}${On leave of absence from Laboratoire d'Annecy-le-Vieux de
Physique Th{\'e}orique  LAPTH, B.P. 110, F-74941 Annecy-le-Vieux et
l'Universit{\'e} de Savoie}}
}\\
\vskip 0.4truecm
%\addresses
$^{*}$ {\it Max-Planck-Institut f\"ur Gravitationsphysik,
Albert-Einstein-Institut, \\
Am M\"uhlenberg 1, D-14476 Golm, Germany}\\
\vskip .2truecm $^{\dagger}$ {\it CERN Theoretical Division, CH-1211 Geneva 23,
Switzerland
} \\
\end{center}

\vskip 1truecm \Large
%\noindent
\centerline{\bf Abstract} \normalsize We study the structure of the four-point
correlation function of the lowest-dimension 1/2 BPS operators (stress-tensor
multiplets) in the (2,0) six-dimensional theory.\\ We first discuss the
superconformal Ward identities and the group-theoretical restrictions on the
corresponding OPE. We show that the general solution of the Ward identities is
expressed in terms of a single function of the two conformal cross-ratios
(``prepotential").\\ Using the maximally extended gauged seven-dimensional
supergravity, we then compute the four-point amplitude in the supergravity
approximation and identify the corresponding prepotential. We analyze the
leading terms in the OPE by performing a conformal partial wave expansion and
show that they are in agreement with the non-renormalization theorems following
from representation theory.\\
The investigation of the (2,0) theory is carried out in close parallel with the
familiar four-dimensional ${\cal N}=4$ super-Yang-Mills theory.
\newpage
\setcounter{page}{1}\setcounter{footnote}{0}

\section{Introduction}  \label{sect0}

One theory of particular interest which has emerged on the scene of the AdS/CFT
duality \cite{M}-\cite{W} is the superconformal six-dimensional theory of
$(2,0)$ self-dual tensor multiplets. It is conjectured to describe the
world-volume fluctuations of the M-theory five-branes and this explains the
special r\^ole the $(2,0)$ theory might play in the possible formulations of
M-theory \cite{Witten}.

Our actual knowledge of the $(2,0)$ theory is very limited, primarily due to
the lack of a field theory formulation. This theory does not allow a
dimensionless coupling and hence a perturbative approach. This is quite
different from the superconformal ${\cal N}=4$ Yang-Mills theory in four
dimensions, where the coupling  is merely a free parameter. On the other hand,
the conformal superalgebras in $d=4$ and $d=6$ and their unitary irreducible
representations (UIR) have a very similar structure. Moreover, both ${\cal
N}=4$ SYM with a gauge group SU(N) and the (2,0) theory possess well-defined
supergravity duals: In the large $N$ limit and for a large t'Hooft coupling the
former  is dual to type IIB ten-dimensional supergravity compactified on
$AdS_5\times S^5$, while the latter is dual to eleven-dimensional supergravity
compactified on $AdS_7\times S^4$, provided the number of tensor multiplets
grows like $N^3$. Thus, representation theory confronts us with the problem to
understand what precisely makes these two theories so different and whether
they have something in common. The aim of the present paper is to provide a
partial answer to this question. At the same time, we put the AdS/CFT
correspondence to another non-trivial test.

We exploit two different but complementary approaches. The first is to study
the operator product expansion (OPE) of two stress-tensor multiplets. These are
the simplest non-trivial examples of the so-called 1/2 BPS operators. Their
conformal dimension is protected from quantum corrections by conformal
supersymmetry, but their OPE has a rich spectrum of both protected and
unprotected multiplets. In Section \ref{sect1.1} we recall the known facts
about this OPE both in $d=4$ and in $d=6$ and make a detailed comparison of the
OPE spectra of the two theories. In particular, we point out the different
realization of what one may call Konishi-like multiplets. In $d=4$ these are
represented by operators bilinear in the fundamental fields which have
canonical dimension and satisfy conservation conditions in the free theory, but
develop anomalous dimension in the presence of interaction. This is related to
the fact that the corresponding superconformal UIRs lie at the unitarity bound
of the {\it continuous} series of representations. At the same time, other
bilinear operators, also at the unitarity bound of the continuous series,
remain protected. This is due to the kinematics of the OPE, i.e., to the
properties of the three-point functions which these operators may form with the
two 1/2 BPS operators.

In $d=6$ the picture is quite different. There the operators at the unitarity
bound of the continuous series of UIRs are trilinear and cannot appear in the
OPE. The closest analogs of the Konishi-like bilinear operators belong to a
{\it discrete} series of UIRs with quantized dimension. Thus, they are
automatically protected by unitarity. The lowest-dimension unprotected
multiplets in this OPE correspond to UIRs lying above the unitarity bound of
the continuous series and are realized by quadrilinear operators.

The second approach is based on the superconformal Ward identities for the
four-point function of stress-tensor multiplets. In $d=4$ they are known to
restrict the freedom in the amplitude to just two functions of conformally
invariant variables, one depending on two such variables and the other on one
variable. An additional, dynamical mechanism which generates the quantum
corrections to the amplitude by insertion of the SYM action, fixes the function
of one variable at its free theory value (``partial non-renormalization"). Once
again, the situation changes in $d=6$. In Sections \ref{sect1.2} and
\ref{sect2} we show that the analogous Ward identities are solved in terms of a
single function of two variables, which we call ``prepotential". In other
words, in $d=6$ the ``partial non-renormalization" is purely a kinematical
effect. We consider this new phenomenon as an indication that there exists no
smooth interpolation between the two fixed conformal points, the free one and
the dual of the supergravity theory.

In the absence of a perturbative formulation of the (2,0) theory one can only
test the above general predictions via the AdS/CFT correspondence. In Sections
\ref{sect1.3} and \ref{sect3} we use the maximally extended gauged
seven-dimensional supergravity to derive the corresponding four-point amplitude
and verify that it satisfies the constraints found in Section \ref{sect2}. We
provide an explicit simple formula for the six-dimensional gravity-induced
prepotential and show how it is related to its four-dimensional analogue.
Finally, in Section \ref{sect4} we analyze the leading terms in the conformal
partial wave expansion of the supergravity four-point amplitude and show that
they are in complete agreement with the OPE structure discussed in Section
\ref{sect1.1}. Some technical details are gathered in the appendices.

\section{Overview and summary of the results}\label{sect1}

\subsection{OPE of stress-tensor multiplets} \label{sect1.1}

One can learn a lot both about the kinematics and the dynamics of a
(super)conformal theory by studying the OPE of various operators. In the
context of the AdS/CFT correspondence the so-called 1/2 BPS short operators
\footnote{In the AdS/CFT literature the BPS operators are often called Chiral
Primary Operators (CPO). This name does not seem very adequate in the (2,0)
theory, where all spinors are chiral.} are of particular interest since, on the
one hand, their conformal dimension is quantized (``protected") in the CFT, and
on the other hand, they can be identified with the Kaluza-Klein excitations in
the AdS supergravity spectrum \cite{W,andfer}. They correspond to states which
are annihilated by half of the supercharges. The simplest example of a 1/2 BPS
operator is the stress-tensor multiplet ${\cal O}^I$ whose lowest component is
a scalar of dimension $\ell=d-2$ belonging to the vector representation of the
R symmetry group SO(6) $\sim$ SU(4) or SO(5) $\sim$ USp(4) in the cases $d=4$
or $d=6$, respectively.

Before discussing the OPE of 1/2 BPS operators, it is useful to recall some
known facts \cite{dp} about the UIRs of the $d=4$
${\cal N}=4$ and the $d=6$ (2,0) superconformal algebras PSU(2,2/4) and
OSp($8^*$/4), correspondingly. They are labeled by the quantum numbers of the
lowest-weight state ${\cal D}(\ell;J_1,J_2;a_1,a_2,a_3)$ (for PSU(2,2/4)) or
${\cal D}(\ell;J_1,J_2,J_3;a_1,a_2)$ (for OSp($8^*$/4)). Here $\ell$ is the
conformal dimension, $J_i$ label the Lorentz group SO(3,1) $\sim$ SL($2,\mathbb
C$) or SO(5,1) $\sim$ SU$^*$(4) irrep, and $[a_i]$ are the Dynkin labels of the
SU(4) or USp(4) irrep, correspondingly. We will be interested in superconformal
UIRs which can appear in the OPE of two 1/2 BPS operators; since the latter are
Lorentz scalars, the former must be vector-like with ``spin" $s$, i.e., with
$J_1=J_2=s/2$ for $d=4$ and $J_1=J_3=0, \ J_2=s$ for $d=6$. Below we list the
relevant UIRs: \footnote{Series B of PSU(2,2/4) contains chiral superfields
with $J_1J_2=0$, so it is not relevant here.}
 \begin{eqnarray}  \nonumber
 &\mbox{PSU(2,2/4),}& J_1=J_2=s/2\, : \\
&{\rm A)}&~~~~~~~\ell\geq 2 + s + a_1 + a_2 + a_3\, , \nonumber \\
&{\rm C)}&~~~~~~~\ell=a_1 + a_2 + a_3 \, ,~~~~~s=0\, ; \label{ser4}
\end{eqnarray}
 \begin{eqnarray}  \nonumber
 &\mbox{OSp($8^*$/4),}& J_1=J_3=0, \ J_2=s\, : \\
&{\rm A)}&~~~~~~~\ell\geq 6+s+2(a_1+a_2)\, , \nonumber \\
\nonumber
&{\rm B)}&~~~~~~~\ell=4+s+2(a_1+a_2) \, ,  \\
\nonumber
&{\rm C)}&~~~~~~~\ell=2+2(a_1+a_2) \, ,~~~~~s=0\, , \\
\label{ser6} &{\rm D)}&~~~~~~~\ell=2(a_1+a_2) \, ,~~~~~s=0\, .
\end{eqnarray}
In both cases series A is continuous whereas B, C and D are isolated and
contain operators with ``quantized" conformal dimension. Series C in $d=4$ and
D in $d=6$ correspond to the BPS states. In particular, if only $a_2 \neq 0$ we
obtain 1/2 BPS states.  The $d=4$ and $d=6$ stress-tensors belong to the 1/2
BPS multiplets ${\cal D}(2;00;020)$ and ${\cal D}(4;000;02)$, respectively.

The OPE of two 1/2 BPS operators is restricted by the kinematics in the sense
that it can contain only the operators whose quantum numbers ${\cal
D}(\ell;J_m;a_n)$ allow them to form a non-vanishing three-point function with
the two 1/2 BPS operators \cite{AES}. The case of interest for us is the OPE of
two stress-tensor multiplets. Then the allowed R symmetry irreps are obtained
by decomposing the tensor products
\begin{eqnarray}\label{tp4}
 {\rm SU(4):} && [020]\times[020]=[040]_{0}+[121]_{0}+[202]_{0}
+[020]_{1}+[101]_{1}+[000]_{2}\,; \\
 {\rm USp(4):} && [02]\times[02]=[04]_{0}+[40]_{0}+[22]_{0}
 +[02]_{1}+[20]_{1}+[00]_{2}\,,\label{tp6}
\end{eqnarray}
where the subscript indicates the value of the integer $k = 2 - \frac{1}{2}\sum
a_i\, .$ In ref. \cite{AES,ES,EFS1,HH1,HH2} it was shown that the
existence of a non-vanishing three-point function  for $k=0,1$ implies certain
selection rules. In particular, the dimension of the operators appearing in the
OPE becomes quantized. No such selection rules are found for $k=2$.  More
specifically, in terms of the classifications (\ref{ser4}) and (\ref{ser6}) the
picture is as follows.

${\mathbf {k=0:}}$\\ The three irreps with $k=0$ in (\ref{tp4}) and (\ref{tp6})
belong to the series C ($d=4$) and D ($d=6$), and therefore they are 1/2 or 1/4
BPS states. The corresponding operators are ``protected", i.e., their conformal
dimension $\ell = 2(d-2)$ cannot be modified by the interaction.

$\mathbf {k=1:}$\\ In this case all the operators are protected as well.
However, this time, besides 1/2 or 1/4 BPS short operators of dimension
$\ell=d-2$, we encounter a new species of protected operators, the so-called
``semishort" operators (see, e.g., \cite{Ferrara:2000eb}). They have different
interpretations in four and six dimensions.

In $d=4$ the semishort operators with spin $s\geq0$ lie at the unitarity bound
$\ell = s + 4$ of the continuous series A. They satisfy conservation-like
conditions in superspace (which imply the existence of conserved component
tensors in the multiplet). For this reason they may also be called
``current-like". We stress that in $d=4$ the semishort operators are {\it a
priori} not protected, since their dimension can be continuously varied above
the unitarity bound. However, the careful analysis of the corresponding
three-point function \cite{AES,ES,EFS1,HH1,HH2} shows that the kinematics of
the particular OPE under consideration protects the dimension of the semishort
operators with $k=1$, so that they remain at the unitarity bound even in the
presence of interaction. A well-known example of a {\it protected semishort}
operator is the so-called ${\cal O}^4_{20}$ corresponding to the UIR ${\cal
D}(4;00;020)$, first discovered in refs. \cite{AFP,AFP1}.

In $d=6$ the semishort operators with $k=1$ correspond to UIRs from the
isolated series B with quantized dimension $\ell=s+8$ \cite{EFS1,EFS2}. Since
their conformal dimension cannot be continuously modified, they are
automatically protected by unitarity. In this respect they resemble the BPS
short operators which belong to an isolated series of UIRs as well. Note that
the existence of an isolated series of semishort operators is specific to the
six-dimensional superconformal algebras OSp($8^*/2{\cal N}$).

${\mathbf {k=2:}}$\\ This is the most interesting case since only it involves
{\it unprotected} operators. As can be seen from (\ref{tp4}) and (\ref{tp6}),
these are R symmetry singlets. Here the analysis of the three-point functions
produces no further selection rules. Still, a particular type of operators can
be singled out. Again, the situation is different in four and in six
dimensions.

In $d=4$ the operators with $k=2$ lying at the unitarity bound $\ell=s+2$ have
twist $\ell-s=2$. So, they correspond to bilinears made out of the free ${\cal
N}=4$ SYM field-strength superfields. Still in the free case, these bilinears
satisfy conservation conditions which make them semishort. However, this
conservation does not reflect any symmetry of the interacting theory, therefore
such operators develop anomalous dimension and drift away from the unitarity
bound. So, the semishort operators with $k=2$ are {\it unprotected}. The
best-known example of this type is the Konishi multiplet (a singlet scalar of
dimension 2), but there exists an infinite series of similar operators with
spin which we call Konishi-like. Their anomalous dimension at one loop has been
calculated in ref. \cite{Anselmi,AFP1,AEPS,Bianchi,DO}.

It should be pointed out that the Konishi-like operators are not present in the
OPE derived from gauged ${\cal N}=8$ supergravity. This was demonstrated in
ref. \cite{AFP} by analyzing the supergravity four-point function of 1/2 BPS
operators found in ref. \cite{AF}. A common lore to explain their absence in
the strongly coupled ${\cal N}=4$ theory is to say that they develop large
anomalous dimension $\ell \sim (g^2_{YM}N)^{1/4}$ as the t'Hooft coupling
$g^2_{YM}N$ tends to infinity, and thus they drop out of the spectrum. Note
that the peculiar asymptotic behavior $(g^2_{YM}N)^{1/4}$ has not yet been
obtained by field theory means and it remains a prediction of string theory.

The six-dimensional case is rather different. Here the multiplets
with $k=2$ lying at the unitarity bound $\ell=6+s$ of the
continuous series A have twist $6$, so in the free theory they
could be realized only by trilinear operators. Such operators
cannot appear in the OPE of two stress-tensor multiplets
(bilinears). Therefore we should look for analogs of the
Konishi-like multiplets among the bilinear composites with $k=2$.
In $d=6$ they should have twist $4$ and we see that they can only
appear in the isolated series B. Thus, we may say that in $d=6$
the Konishi-like semishort multiplets are protected by unitarity.

Being protected operators in $d=6$, the Konishi-like multiplets are not
expected to appear in the supergravity-induced OPE. This can be anticipated on
the general grounds of the AdS/CFT correspondence, because there is no field in
the spectrum of the corresponding supergravity theory dual to any of these
currents.  In Section \ref{sect3}, by using gauged seven-dimensional ${\cal
N}=4$ supergravity \cite{PPN}, we compute the four-point amplitude of the
lowest dimension 1/2 BPS operators in the $(2,0)$ theory. Subsequently, in
Section \ref{sect4} we indeed demonstrate the absence of the Konishi-type
currents in the supergravity-induced OPE.  On the other hand, the Konishi-like
multiplets are present in the free OPE of two 1/2 BPS operators. In our
opinion, the fact that they drop out of the spectrum of the interacting theory
clearly demonstrates the absence of a superconformal theory that could smoothly
interpolate between the free CFT and the CFT dual to the eleven-dimensional
supergravity on the $AdS_7\times S^4$ background.

Note that the $d=6$ OPE under consideration does not contain
operators from series C. Indeed, since in the $k=2$ channel
$a_1=a_2=0$, they should have the dimension of the fundamental
field $\ell=2$.

\subsection{Four-point function of stress-tensor multiplets} \label{sect1.2}

The complete, i.e., both kinematical and dynamical information about the OPE of
two stress-tensor multiplets is encoded in their four-point correlation
function. We have already seen that the kinematics (or, in other terms,
superconformal representation theory) strongly restricts the content of the
OPE. We should expect to see the implications of these restrictions on the
four-point amplitude. The easiest and most economic way to do this is to use
the superconformal Ward identities. Below we summarize the already known
results about this four-point amplitude in $d=4$
\cite{hw1,hw2,EHPSW,EPSS,DO2,DO} and compare them to our new results in the
six-dimensional case.

We would like to stress that the four-point amplitude of 1/2 BPS short
multiplets that we consider is rather special in the sense that superconformal
symmetry is powerful enough to restore the complete superspace dependence
solely from the knowledge of the lowest ($\theta=0$) component of the
amplitude. Indeed, the 1/2 BPS short superfields depend on half of the
Grassmann variables. Thus, a four-point function of this type depends on
$4\times (1/2)=2$ full sets of odd variables. At the same time, the
superconformal algebra has two sets of odd shift-like generators (Q and S
supersymmetry). This leaves no room for nilpotent superconformal invariants
made out of the odd variables and thus the $\theta$ expansion is completely
fixed.

The lowest component of this amplitude corresponds to the correlator of four
scalar operators ${\cal O}^{I}$ of dimension $\ell=d-2$ in the vector
representation of the R symmetry group SO(6) (for $d=4$) or SO(5) (for $d=6$):
\begin{eqnarray}
\label{Amp} && \hskip-5mm \langle {\cal O}^{I_1}(x_1)\ldots {\cal
O}^{I_4}(x_4)\rangle =
 \nonumber\\
 &&a_1(s,t)\frac{\d^{I_1I_2}\d^{I_3I_4}}{(x_{12}^2x_{34}^2)^{d-2}}
+a_2(s,t)\frac{\d^{I_1I_3}\d^{I_2I_4}}{(x_{13}^2x_{24}^2)^{d-2}}
+a_3(s,t)\frac{\d^{I_1I_4}\d^{I_2I_3}}{(x_{14}^2x_{23}^2)^{d-2}} \\
\nonumber
 && +\,
 b_1(s,t)\frac{C^{I_1I_3I_2I_4}}{(x_{13}^2x_{14}^2x_{23}^2x_{24}^2){\frac{d-2}{2}}}
+b_2(s,t)\frac{C^{I_1I_2I_3I_4}}{(x_{12}^2x_{14}^2x_{23}^2x_{34}^2){\frac{d-2}{2}}}
+b_3(s,t)\frac{C^{I_1I_2I_4I_3}}{(x_{12}^2x_{13}^2x_{24}^2x_{34}^2){\frac{d-2}{2}}}
\, .
\end{eqnarray}
Here $s,t$ are the conformal cross-ratios
\begin{equation*}
 s = {x^2_{12}x^2_{34}\over x^2_{13}x^2_{24}}\;, \qquad
t = {x^2_{14}x^2_{23}\over x^2_{13}x^2_{24}}\;.
\end{equation*}
The six tensor structures $\d^{I_1I_2}\d^{I_3I_4}$, $C^{I_1I_2I_3I_4}$ (and
permutations) are invariant tensors of SO(6) (or SO(5)) and are related to the
six channels in the OPE (\ref{tp4}) (or (\ref{tp6})). In general, we define the
invariant tensors $C^{I_1\ldots I_n}$ as $\mbox{tr}(C^{I_1}\ldots C^{I_n})$,
where the matrices $C_{ij}^I$, which are symmetric and traceless in their lower
indices, realize a basis of the corresponding vector representation of the R
symmetry group.

Among the six coefficient functions in (\ref{Amp}) only two are independent,
for example, $a_1(s,t)$ and $b_2(s,t)$. The others are obtained from the
crossing symmetry relations
\begin{eqnarray}
  &&a_1(s,t) =  a_3(t,s) =  a_1(s/t,1/t)\nonumber\\
  &&a_2(s,t) = a_2(t,s) = a_3(s/t,1/t)\nonumber\\
  &&b_1(s,t) = b_3(t,s) =  b_1(s/t,1/t) \label{permut12}\\
  &&b_2(s,t) = b_2(t,s) =  b_3(s/t,1/t)  \nonumber
\end{eqnarray}

The amplitude (\ref{Amp}) must obey superconformal Ward identities
which follow from the 1/2 BPS nature of the supermultiplets. In
the four-dimensional case they take the form of two first-order
PDEs for the independent coefficient functions \cite{EHPSW}:

$\qquad \mbox{\underline{$d=4$ Ward identities:}}$
\begin{eqnarray}
  %&&\mbox{\underline{$d=4$ Ward identities:}} \nonumber\\
  &&\pa_t b_2=\frac{s}{t}\pa_s
a_3-\pa_s a_1-\frac{s+t-1}{s}\pa_t a_1  \nonumber\\
  &&\pa_s b_2=\frac{t}{s}\pa_t a_1 -\pa_t a_3 - \frac{s+t-1}{t}\pa_s a_3  \label{d4con}
\end{eqnarray}
In six dimensions the corresponding equations look very similar (see Section
\ref{sect2} for the derivation):

$\qquad\mbox{\underline{$d=6$ Ward identities:}}$
\begin{eqnarray}
  %&&\mbox{\underline{$d=6$ Ward identities:}} \nonumber\\
  &&\pa_t b_2=\frac{s^2}{t^2}\pa_s
a_3-\frac{t}{s}\pa_s a_1-\frac{t(s+t-1)}{s^2}\pa_t a_1  \nonumber\\
  &&\pa_s b_2=\frac{t^2}{s^2}\pa_t a_1 -\frac{s}{t}\pa_t a_3 -
\frac{s(s+t-1)}{t^2}\pa_s a_3
  \label{d6con}
\end{eqnarray}
However, the small change in the coefficients from eqs. (\ref{d4con}) to eqs.
(\ref{d6con}) results in an important difference when it comes to their general
solution.

In ref. \cite{EPSS} it was found that the general solution \footnote{We recall
some details in Appendix 1.} of the $d=4$ Ward identities (\ref{d4con}) is
parametrized by {\it two independent functions}, one of two variables and the
other of a single variable. \footnote{It should be mentioned that a similar
picture has been observed many years ago by {}Fradkin, Palchik and Zaikin
\cite{FPZ}. They have studied the conformal correlator of one conserved current
with three scalar operators. By imposing the Ward identity for the current they
have found a differential equation whose solution has exactly the same
functional freedom. However, the difference between their study and ours is in
the fact that the functions determining their correlator are obtained as
derivatives of our coefficients $a_i$ and $b_i$ (when reconstructing the
corresponding component of our superamplitude starting from the lowest one).
Thus, the solution of our constraints lies one level deeper than that of ref.
\cite{FPZ}.} \footnote{For a recent discussion in which crossing symmetry is
not imposed see ref. \cite{DO}.} It was further shown in ref. \cite{EPSS} that
the function of one variable can be set to its free-theory value by evoking a
dynamical mechanism. It consists in employing Intriligator's insertion
procedure \cite{I} which gives the quantum (interacting) part of the amplitude
as the result of the insertion of the SYM action into it. Thus, combining
kinematics with dynamics, the full solution of the $d=4$ superconformal Ward
identities is reduced to
\begin{eqnarray}
\nonumber
a_1(s,t)&=&A+s {F}(s,t) \, ,\\
\nonumber b_2(s,t)&=&B+(1-s-t){F}(s,t) \, ,
\end{eqnarray}
where $A$ and $B$ are constants determined from the free ${\cal N}=4$ theory.

The non-trivial part of the amplitude is therefore encoded in the
{\it single function of two variables} ${F}(s,t)$ satisfying the
crossing-symmetry conditions
\begin{equation}\label{crosym0}
  { F}(s,t) = { F}(t,s) = 1/t\, { F}(s/t,1/t) \, .
\end{equation}
It includes all (non-)perturbative corrections to the free field amplitude.
This prediction of the superconformal Ward identities and of the dynamical
insertion procedure about the form of the amplitude, called ``partial
non-renormalization" in ref. \cite{EPSS}, has been confirmed by all the
available perturbative \cite{gps}, instanton \cite{BGKR}
and strong coupling \cite{AF} results.

In six dimensions the solution of the Ward identities (\ref{d6con}) is directly
given in terms of {\it one unconstrained function of two variables} (see
Section \ref{sect2}):
\begin{eqnarray}
\nonumber
a_1(s,t)&=&A+s^4\D  \Biggl(\frac{1}{\lambda^3}t{\cal F}(s,t)\Biggr) \, ,\\
\nonumber b_2(s,t)&=&B+s^2t^2 \D \Biggl(\frac{1}{\l^3}(1-s-t) {\cal F}(s,t)
\Biggr)\, ,
\end{eqnarray}
where $A$ and $B$ are additive integration constants, $\D$ is a second-order
differential operator,
\begin{eqnarray} \label{op}
\Delta=s\pa_{ss}+t\pa_{tt}+(s+t-1)\pa_{st}+3\pa_s+3\pa_t \, ,
\end{eqnarray}
and $\lambda=\sqrt{(s+t-1)^2-4st}$ is its discriminant. Here the function
${\cal F}(s,t)$ satisfies the same crossing-symmetry relations as its
four-dimensional analogue (recall (\ref{crosym0})):
\begin{equation}\label{crosym}
  {\cal F}(s,t) = {\cal F}(t,s) = 1/t\, {\cal F}(s/t,1/t) \, .
\end{equation}
Again, it encodes the dynamics of the theory and, in particular, it comprises
all M-theory corrections to the leading supergravity result. However, in $d=4$
this function itself is a coefficient function of the amplitude, whereas in
$d=6$ it plays the r\^ole of a {\it prepotential} in the sense that all the
coefficients can be obtained from it by applying derivatives. It would be very
interesting to find out whether this prepotential has a deeper origin.

We would like to underline once more the important difference between the four-
and six-dimensional cases. In $d=4$ one can reduce the freedom in the amplitude
to just {\it one unconstrained function} by combining kinematics (the
superconformal Ward identities) with dynamics (the insertion formula). The
latter relies on the existence of a certain nilpotent superconformal five-point
covariant with rather special properties \cite{Eden:1999gh,EPSS}. Our attempts
to find a similar construction in $d=6$ were unsuccessful. This again points at
the absence of a Lagrangian formulation of the six-dimensional theory. However,
now we see that in $d=6$ the kinematics (superconformal symmetry) alone leaves
{\it exactly the same freedom}, the single function ${\cal F}(s,t)$.

Our final remark concerns an alternative explanation of the r\^ole of the
function of one variable in the four-dimensional amplitude, recently discussed
by Dolan and Osborn \cite{DO}. They relate this function to the possible
exchanges only of protected operators in the OPE (the first five channels in
the decomposition (\ref{tp4})). Indeed, it is easy to show that Intriligator's
insertion procedure forbids such exchanges \cite{HH2}, and so it is natural to
expect that this fixes the function at its free-theory value. However, in six
dimensions similar protected channels exist, but the insertion procedure cannot
be applied. Still, we do not find a function of one variable here. It would be
interesting to understand this phenomenon from the OPE point of view advocated
in ref. \cite{DO}. One might speculate about the different behavior of the
protected semishort operators in $d=4$ which lie at the boundary of the
continuous series A, and of those in $d=6$ which belong to the isolated series
B.

\subsection{Obtaining the prepotential \mbox{${\cal F}$} from gauged supergravity} \label{sect1.3}

Since no field-theory formulation of the interacting $(2,0)$ six-dimensional
theory is available, the way to check the general predictions we have found
here is to compute the amplitude via the AdS/CFT correspondence and to try to
identify the prepotential ${\cal F}$. We perform this program in Section
\ref{sect3} by using the gauged seven-dimensional ${\cal N}=4$ supergravity and
find a perfect agreement. In particular, we show that the supergravity
four-point amplitude of the 1/2 BPS operators is generated by the following
very simple prepotential
\begin{eqnarray} \label{prep13} {\cal
F}(s,t)=\frac{240}{N^3}\, \frac{\l^3}{st^2}\, \bar{D}_{7333}(s,t) \,,
\end{eqnarray}
together with the integration constants \bea A=1\, , \qquad B=\frac{1}{N^3} \,
. \eea The conformally covariant  functions $\bar{D}_{\D_1\D_2\D_3\D_4}(s,t)$
are defined in Appendix 2.

It is interesting to compare this supergravity-induced solution with the theory
of $\eta$ free (2,0) tensor multiplets. For this theory the prepotential ${\cal
F}$ vanishes, while the constants $A$, $B$ equal
\begin{eqnarray} A=1\, , \qquad B=\frac{4}{\eta} \, . \end{eqnarray}
If one would try to view the supergravity solution ${\cal F}$ as being obtained
from the free one ${\cal F}=0$ by some smooth deformation, then one should
obviously set $\eta=4N^3$. The factor $4N^3$ was found in ref. \cite{K}
by studying the absorption rate of longitudinally polarized gravitons by M5
branes. The same factor appears as the universal coefficient between the free
and the AdS two- and three-point correlators of the stress tensor \cite{BFT1},
as well as between the free and the AdS type B conformal anomaly
\cite{BFT2}. Since all the non-trivial dynamics is encoded in the
prepotential, we see that $4N^3$ is just what is needed to match the
integration constants of the free and of the supergravity-induced four-point
amplitudes. In Section \ref{sect3} we discuss however that the existence of a
smooth superconformal deformation from the free to the supergravity theory
appears to be in conflict with unitarity. \footnote{One way to smoothly connect
the free and the supergravity amplitudes is to multiply the supergravity
prepotential by a function $f(g)$ of some ``coupling constant" $g$, such that
$f(0)=0$ and $f(1)=1$. This is a trivial possibility that would explain the
decoupling of the Konishi-like multiplets from the supergravity OPE by the
vanishing of the corresponding OPE coefficients. In what follows we discard
such deformations.}

It is instructive to compare the six-dimensional prepotential (\ref{prep13})
with the ``potential" which generates the strongly-coupled $d=4$ ${\cal N}=4$
amplitude found via the AdS/CFT correspondence \cite{AF,EPSS}: \footnote{See
also Appendix D \ of ref. \cite{DO}, where a similar formula for $F(s,t)$ was
extracted from the results of \cite{AF}.} \bea \label{pot4}
F(s,t)=-\frac{24}{N^2}\, \frac{1}{t}\, \bar{D}_{4222}(s,t)\, ,\qquad A=1\,
,\qquad B=\frac{4}{N^2} \, . \eea Since differentiating a $\bar{D}$-function
with respect to $s$ or $t$ amounts to raising by unity the values of two of its
indices, we see that the six-dimensional prepotential is obtained from the
four-dimensional potential by {\it dressing} it with a certain third-order
differential operator (see Section 4). It would be interesting to find out if
this operator has some intrinsic meaning.

Finally, we observe that we might construct infinite towers of superconformal
four-point amplitudes both in the $d=4$ and $d=6$ theories as follows: \bea
\label{newampl} F(s,t)\sim \frac{1}{t}\, \bar{D}_{3\D-2,\D,\D,\D}(s,t)\,
,\qquad {\cal F}(s,t)\sim \frac{\l^3}{st^2}\, \bar{D}_{3\D-2,\D,\D,\D}(s,t)\, ,
\eea where $\D=1,2,\ldots .$ These functions are symmetric and satisfy
(\ref{crosym}) as a consequence of the corresponding symmetry properties
(\ref{tr1}) and (\ref{tr2}) of the $\bar{D}$-functions. One could ask the
question whether any of the amplitudes (\ref{newampl}) (or their linear
combinations), other then (\ref{prep13}) and (\ref{pot4}), together with some
appropriate integration constants $A$ and $B$, has an OPE free from
Konishi-like multiplets. If not, this might explain the distinguished r\^ole of
the amplitudes (\ref{prep13}) and (\ref{pot4}).

\section{General structure of the four-point amplitude}\label{sect2}

We begin this section by recalling, just in a few words, the procedure which
leads to the $d=4$ Ward identities (\ref{d4con}). The origin of these
constraints can be traced back to the fact that the ${\cal N}=4$ SYM
stress-tensor multiplet is 1/2 BPS short. The natural framework for describing
BPS shortness is harmonic superspace \cite{GIKOS}. The constraints
(\ref{d4con}) can be derived directly in ${\cal N}=4$ harmonic superspace
\cite{HH}, but it is easier to do this using its ${\cal N}=2$ version (both
methods are explained in detail in ref. \cite{EHPSW}; for a recent rederivation
of the same constraints without using harmonic superspace see ref. \cite{DO}).
The main point is that when reconstructing the full harmonic superspace
dependence of the four-point amplitude starting from its lowest component
(\ref{Amp}), one encounters harmonic singularities already at the level
$(\theta)^2$. Their absence, i.e., the requirement of harmonic analyticity, is
equivalent to imposing irreducibility under the R symmetry group. It is
precisely this requirement which leads to the constraints (\ref{d4con}). Once
these constraints have been imposed, it can be shown that no new harmonic
singularities appear at the higher levels of the $\theta$ expansion of the
amplitude.

In $d=6$ one could go through exactly the same steps in order to obtain the new
constraints (\ref{d6con}). However, there is a much faster way which consists
in simply adapting the $d=4$ constraints (\ref{d4con}) to the case $d=6$. The
key observation is that the coefficient functions $a_1,a_3,b_2$ appear in
(\ref{d4con}) only through their first-order derivatives. The origin of these
derivatives is in the completion of the conformal invariant $s$ to a full
superconformal invariant $\hat s = s + \mbox{$\theta$-terms}$ (and similarly
for $t$). Expanding, e.g., $a_1(\hat s, \hat t)$ up to the level $(\theta)^2$
gives rise to the terms $\pa_s a_1, \pa_t a_1$.  It is not hard to show that
$\hat s, \hat t$ are exactly the same both in $d=4$ and in $d=6$. Next, the
rational coefficients in (\ref{d4con}) originate from the ``propagator" factors
$1/(x^2_{12}x^2_{34})^{d-2}$, etc. in (\ref{Amp}). These differ in $d=4$ and
$d=6$, as can be seen most clearly by pulling out one of them in front of the
amplitude (only the relevant terms are shown):
\begin{equation*}
%\label{pullout}
%\nonumber
  \frac{1}{(x_{12}^2x_{34}^2)^{d-2}} \left[a_1(s,t)\, \d^{I_1I_2}\d^{I_3I_4} +
  \left(\frac{s}{t} \right)^{d-2}a_3(s,t)\, \d^{I_1I_4}\d^{I_2I_3} +
   \left(\frac{s}{t} \right)^{\frac{d-2}{2}}b_2(s,t)\, C^{I_1I_2I_3I_4} + \cdots \right]
\end{equation*}
Now, the completion of these propagator factors to full superconformal
covariants does not affect the derivative terms in (\ref{d4con}). Therefore, in
order to pass from $d=4$ to $d=6$ it is sufficient to just redefine the
coefficient functions as follows:
\begin{equation}\label{redef}
 a_1^{d=6} \ \rightarrow\  a_1^{d=4}\,, \qquad  a_3^{d=6}  \ \rightarrow\  \frac{s^2}{t^2}
\, a_3^{d=4}\,, \qquad b_2^{d=6}  \ \rightarrow\  \frac{s}{t}\, b_2^{d=4}\,.
\end{equation}
This redefinition should be done in (\ref{d4con}) so that the derivatives do
not act on the factors in (\ref{redef}). The result is precisely the
constraints (\ref{d6con}).

We remark that due to the crossing symmetry relations (recall (\ref{permut12}))
\begin{eqnarray} a_3(t,s)=a_1(s,t)\,, ~~~~~~~b_2(s,t)=b_2(t,s) \end{eqnarray}
the second equation both in (\ref{d4con}) and in (\ref{d6con})  is
automatically satisfied.

{}From here on we concentrate on the solution of the $d=6$ constraints
(\ref{d6con}). The integrability condition for this system of PDEs is
\begin{eqnarray}
\label{2ndodcon} t\Delta\left(\frac{a_1}{s^2}\right)=
s\Delta\left(\frac{a_3}{t^2}\right) \end{eqnarray} where $\D$ was defined in
(\ref{op}). To solve this second-order PDE we make the substitution
\begin{eqnarray} \label{sub1} \frac{a_1}{s^2}=\frac{s}{\lambda^3}\, G(s,t)  \,~\,
,~~~ ~~~\frac{a_3}{t^2}=\frac{t}{\lambda^3}\, G(t,s)
\end{eqnarray} and change the variables $s,t$ to the ``normal
coordinates" $x,y$ for the hyperbolic operator $\D$: \bea x=\rho s\, , \qquad
y= \rho t \, , \eea where $\rho=2(1-s-t+\l)^{-1}$ and
$\l=\sqrt{(1-s-t)^2-4st}$.  After this change eq. (\ref{2ndodcon}) becomes
\begin{eqnarray} y\left(\pa_y+x^2\pa_x+(1-xy)\pa_{xy}\right)G(x,y)=
x\left(\pa_x+y^2\pa_y+(1-xy)\pa_{xy}\right)G(y,x) \, . \nonumber \end{eqnarray}
Then we set \begin{eqnarray}
%\label{sub2}
\nonumber G(x,y) = \phi(x,y) + \gamma(x,y) \, ,
\end{eqnarray}
where
\begin{equation}\label{symcon}
  \phi(x,y)= \phi(y,x)\,,\qquad \gamma(x,y)=-\gamma(y,x) \,
\end{equation}
and perform one more change of variables
\begin{eqnarray}
%\label{logvar}
\nonumber \sigma = \ln(xy) \, ,\qquad  \tau=\ln \frac{x}{y}\,.
\end{eqnarray}
In these new variables the symmetry conditions (\ref{symcon}) become
\begin{equation}\label{symcon1}
  \phi(\sigma ,\tau)=\phi(\sigma ,-\tau)\,,\qquad \gamma(\sigma ,\tau)=-\gamma(\sigma ,-\tau)
\end{equation}
and equation (\ref{2ndodcon}) takes the form
\begin{eqnarray}
 \phi_\tau=\g_{\sigma \sigma }-\g_{\tau\tau}-\coth(\sigma /2) \g_\sigma  \ .
\end{eqnarray}

This equation can be integrated to give \begin{eqnarray} \nonumber
\phi=\left(\pa_{\sigma \sigma }-\coth(\sigma /2) \pa_\sigma
\right)\int_{\tau_0}^{\tau}\g(\sigma ,\tau')d\tau'-\g_\tau+c(\sigma ) \, ,
\end{eqnarray} where $c(\sigma )$ is an integration ``constant" depending on
$\sigma $. Let us introduce the function
\begin{eqnarray}{\cal F}(\sigma ,\tau)=-\int_{\tau_0}^{\tau}\g(\sigma ,\tau')d\tau'\, . \label{Integ}
\end{eqnarray} It is even, ${\cal F}(\sigma ,\tau)={\cal F}(\sigma ,-\tau)$ and is supposed to obey the boundary
condition ${\cal F}(\sigma ,\tau_0)=0$, where $\tau_0$ is some arbitrary fixed
point. Without loss of generality $c(\sigma )$ is absorbed into ${\cal F}$,
which results only in a change of the boundary condition for ${\cal F}$. Thus,
\begin{eqnarray} \label{sol} \phi+\gamma=
\left[\pa_{\tau\tau}-\pa_{\sigma \sigma }+\coth(\sigma /2)
\pa_\sigma -\pa_\tau\right]{\cal F} \end{eqnarray} and the function ${\cal F}$
plays the r\^ole of a {\it prepotential}.

Switching back to the original variables $s,t$ we find \begin{eqnarray}
\label{a1}
a_1&=&\frac{s^4t}{\lambda^3}\Biggl[2\pa_t+\frac{s+t-1}{t}\pa_s+s\pa_{ss}+(s+t-1)\pa_{st}+t\pa_{tt}\Biggr]{\cal F}(s,t) \, ,\\
\label{a3}
a_3&=&\frac{t^4s}{\lambda^3}\Biggl[2\pa_s+\frac{s+t-1}{s}\pa_t+s\pa_{ss}+(s+t-1)\pa_{st}+t\pa_{tt}\Biggr]{\cal
F}(s,t) \, , \end{eqnarray} where the prepotential ${\cal F}(s,t)$ is an
arbitrary symmetric function.

Some comments are due here. First of all, it is obvious that $a_1$
and $a_3$ admit the constant solution $a_1=a_3=A$. Computing
$\gamma(\sigma,\tau)$ for this trivial solution and further integrating
over $\tau$ we find the corresponding prepotential ${\cal F}_0(s,t)$:
\begin{eqnarray} {\cal F}_0(s,t)=
\frac{A}{3}\frac{s^3+t^3}{s^3t^3}\l^3 \, . \end{eqnarray} Secondly, the
prepotential is not uniquely defined. The freedom in redefining ${\cal F}$
without changing the amplitude can be easily found by solving the homogenous
equation implied by (\ref{sol}). This is done by separating the variables. The
resulting freedom is
\begin{eqnarray} \label{gauge}{\cal F}(s,t)\to {\cal F}(s,t)+ h(s,t)\, ,\end{eqnarray} where \begin{eqnarray}
h(s,t)=
h_1\left(st\rho^2-\frac{1}{st\rho^2}-2\ln(st\rho^2)\right)+h_2\,
\end{eqnarray} and $h_1$ and $h_2$ are arbitrary constants.

Since the integrability condition (\ref{2ndodcon}) is already satisfied, we can
now integrate, e.g., the first of the equations (\ref{d6con}) for $b_2$. We
obtain
\begin{eqnarray} \label{b2}
b_2(s,t)&=&B-\frac{s^2t^2}{\l^3}(s+t-1)\Biggl[s\pa_{ss}+(s+t-1)\pa_{st}+t\pa_{tt} \\
\nonumber &+&\left(1+\frac{2s}{s+t-1}\right)\pa_s+
\left(1+\frac{2t}{s+t-1}\right)\pa_t\Biggr] {\cal F}(s,t) \,
,\end{eqnarray} where $B$ is a new integration constant. Under the
replacement (\ref{gauge}) the coefficients $a_1$ and $a_3$ and,
therefore, eqs. (\ref{d6con}) remain unchanged. However, the
solution (\ref{b2}) is allowed to pick an additive constant.
Indeed, we find that (\ref{gauge}) leads to the shift $B\to
B+h_1$. Finally, note that the constant $B$ remains unchanged
under the replacement ${\cal F}(s,t)\to {\cal F}(s,t)+{\cal
F}_0(s,t)$.

Now we are in a position to find the implications of the global crossing
symmetry conditions. Under the change $s\to s/t$, $t\to 1/t$ we find
\begin{eqnarray} a_1(s/t, 1/t)=\frac{s^4}{\lambda^3}
\Biggl[s\pa_{ss}+(s+t-1)\pa_{st}+t\pa_{tt}\Biggr]{\cal F}\left(s/t,1/t\right)
\, .
\end{eqnarray} Clearly,  by choosing \begin{eqnarray} \label{cross}
{\cal F}\left(s/t,1/t\right)=t{\cal F}\left(s,t\right)\end{eqnarray} we are
able to satisfy the crossing symmetry relation $a_1(s/t,1/t)=a_1(s,t)$. Note
that neither ${\cal F}_0(s,t)$ nor $h(s,t)$ obey eq. (\ref{cross}). Therefore,
in the following we represent the general solution of the superconformal Ward
identities in the form
\begin{eqnarray} {\cal F}_0(s,t)+{\cal F}(s,t)\, , \end{eqnarray} where ${\cal F}(s,t)$ satisfies
the crossing symmetry relation (\ref{cross}). This requirement also fixes the
freedom (\ref{gauge}) in the prepotential.

Formulae (\ref{a1}), (\ref{a3}) and (\ref{b2}) can be further simplified to
give 
\begin{eqnarray} \nonumber
a_1&=&\frac{s^4}{\lambda^3}\Biggl[s\pa_{ss}+(s+t-1)\pa_{st}+t\pa_{tt}\Biggr]\Biggl(t{\cal F}(s,t)\Biggr) \, ,\\
\label{prom}
a_3&=&\frac{t^4}{\lambda^3}\Biggl[s\pa_{ss}+(s+t-1)\pa_{st}+t\pa_{tt}\Biggr]\Biggl(s{\cal F}(s,t) \Biggr)\, , \\
\nonumber b_2&=&
\frac{s^2t^2}{\l^3}\Biggl[s\pa_{ss}+(s+t-1)\pa_{st}+t\pa_{tt}\Biggr]\Biggl((1-s-t)
{\cal F}(s,t) \Biggr)\, .
\end{eqnarray}
Now we note that the operator $\D$ has the following property. For any function
$w(s,t)$,
\begin{eqnarray}
\nonumber \Delta
\frac{w(s,t)}{\lambda^3}=\frac{1}{\lambda^3}\Biggl[t\pa_{tt}+(s+t-1)\pa_{st}+s\pa_{ss}
\Biggr]w(s,t)\,.
\end{eqnarray}
This allows us to move the factor $1/\l^3$ to the right through the
differential operator in eqs. (\ref{prom}).

We thus obtain the complete solution for the coefficients of the four-point
amplitude in terms of the prepotential ${\cal F}(s,t)$:
\begin{eqnarray}
\nonumber
a_1(s,t)&=&A+s^4\D  \Biggl(\frac{1}{\lambda^3}t{\cal F}(s,t)\Biggr) \, ,\\
\nonumber
a_2(s,t)&=&A+\D  \Biggl(\frac{1}{\lambda^3}st{\cal F}(s,t)\Biggr) \, ,\\
\nonumber
a_3(s,t)&=&A+t^4\D  \Biggl(\frac{1}{\lambda^3}s{\cal F}(s,t) \Biggr)\, , \\
\label{Gamplitude}
b_1(s,t)&=&B+t^2 \D \Biggl(\frac{1}{\l^3}s(s-t-1) {\cal F}(s,t) \Biggr)\, ,\\
\nonumber
b_2(s,t)&=&B+s^2t^2 \D \Biggl(\frac{1}{\l^3}(1-s-t) {\cal F}(s,t) \Biggr)\, ,\\
\nonumber b_3(s,t)&=&B+s^2 \D \Biggl(\frac{1}{\l^3}t(t-s-1) {\cal F}(s,t)
\Biggr)\, . \end{eqnarray} In this form the crossing symmetry relation is most
transparent, given that the operator $\D$ transforms as
\begin{eqnarray} \D_{s/t,1/t}(t^2w(s,t))=t^4\D_{s,t}(w(s,t)) \end{eqnarray} for any
function $w(s,t)$. Finally, note that $\l$ is a symmetric function of $s,t$,
however under $s\to s/t$, $t\to 1/t$ it transforms as $\l \to \l/t$. Thus, if
we redefine ${\cal F}$ as ${\cal F}\to \l^{-3}{\cal F}$, the new function obeys
the crossing symmetry relation ${\cal F}\left(s/t,1/t\right)=t^4{\cal
F}\left(s,t\right)$.

\section{Four-point amplitude from gauged supergravity} \label{sect3}

According to the duality conjecture for the (2,0) theory, in the supergravity
regime the correlation functions of any 1/2 BPS operators and of their
supersymmetry descendants can be computed from eleven-dimensional supergravity
on an $AdS_7\times S^4$ background. In this way many two- and three-point
correlation functions have already been found \cite{AF1}. 
%These include, in particular, the two- and three-point amplitudes of the stress
%tensor \cite{AF1,BFT2}. 
Below we present the first example of a four-point
amplitude of 1/2 BPS operators in this theory and subsequently analyze the
leading terms of the underlying OPE. The operators whose amplitude we are going
to find, have the lowest scaling dimension $\ell=4$ and their dual supergravity
scalars belong to the massless graviton multiplet of the $AdS_7\times S^4$
compactification. Thus, for our present purposes it is enough to consider only
the sector of the theory described by gauged seven-dimensional supergravity.

The gauged seven-dimensional ${\cal N}=4$ supergravity was
constructed in ref. \cite{PPN} by gauging Poincar\'{e}
supergravity. Alternatively, it can be obtained by compactifying
eleven-dimensional supergravity on $AdS_7\times S^4$ with a
further Kaluza-Klein truncation to the massless graviton multiplet
\cite{NVN}.

The bosonic sector of the theory consists of the metric $g_{\m\n}$, fourteen
scalars parametrizing the coset space SL(5,$\mathbb R$)/SO(5)${}_c$, the
SO(5)${}_g$ Yang-Mills gauge fields $A_{\mu}^{IJ}$ and a five-plet of
antisymmetric tensors $S_{\mu\nu\rho}^I$, $I,J=1,\ldots , 5$. The relevant part
of the Lagrangian is (the metric is assumed to have Minkowskian signature)
\begin{eqnarray} e^{-1}{\cal L}=R+\frac{g^2}{8}(T^2-2T_{ij}T^{ij})
-\frac{1}{4}P_{\m \, ij}P^{\m\, ij}
-\frac{1}{2}F_{\m\n}^{IJ}F^{\m\n}_{IJ}\, .
\label{action}\end{eqnarray} Here $F_{\m\n}^{IJ}$ is the field
strength for $A_{\mu}^{IJ}$. To describe the scalar manifold one
introduces the vielbein $(S^{-1})_I^i\in $ SL(5,$\mathbb R$),
where $i=1,\ldots, 5$ is an SO(5)${}_c$ index. Then
$T_i^j=(SS^t)_i^j$ and $T=\Tr(SS^t)$. The kinetic term is given by
the matrix $P_{\mu}$\, : \begin{eqnarray} \nonumber P_{\m
}=S\nabla_{\mu}S^{-1}-gSA_{\m}S^{-1}+
\nabla_{\mu}(S^{-1})^tS^t+g(S^{-1})^tA_{\m}S^{t}\, ,
\end{eqnarray} where $g$ is the Yang-Mills coupling. Note that in
writing eq. (\ref{action}) we omitted the part of the action
depending on the antisymmetric fields since, as can be easily
shown, the latter do not propagate in the AdS exchange graphs
involving four external scalar fields.

To proceed, we choose the following natural parametrization for $S$:
$S=e^{\Lambda}$ where $\Lambda$ is a traceless symmetric $5\times 5$ matrix.
The scalar fields parametrizing $\Lambda$ are dual to the 1/2 BPS operators
${\cal O}^I$ of dimension $\ell=4$ with the index $I$ transforming under the
irrep $[02]$ of the R symmetry group SO(5).

Since we are interested in the four-point amplitude of the operators ${\cal
O}^I$, we decompose the Lagrangian in power series in $\Lambda$ and then
truncate it at the fourth order. The resulting expression reads
\begin{eqnarray} \nonumber e^{-1}{\cal L}&=&R-\Biggl( \nabla_{\mu}\Lambda
\nabla^{\mu}\Lambda + \frac{2}{3}\nabla_{\mu}\Lambda \Lambda^2
\nabla^{\mu}\Lambda
-\frac{2}{3}\nabla_{\mu}\Lambda\Lambda\nabla_{\mu}\Lambda\Lambda+2g\nabla^{\mu}\Lambda
[\Lambda,A_{\mu}]
\Biggr) \\
&+&\frac{g^2}{8}\Biggl(15+4\mbox{tr}\Lambda^2-8\mbox{tr}\Lambda^3+4\left(\mbox{tr}\Lambda^2\right)^2
-\frac{44}{3}\mbox{tr}\Lambda^4 \Biggr) -\frac{1}{2}F_{\m\n}^{IJ}F^{\m\n}_{IJ}
\, .\end{eqnarray} Obviously, to obtain the correct value
$2\lambda=-(d-1)(d-2)=-30$ of the cosmological constant in $d=7$, one has to
set $g^2=16$.

To simplify the resulting expression, we perform the field redefinition
$\Lambda\to \Lambda-\frac{2}{3}\Lambda^3$. It is also convenient to introduce
another parametrization for the matrix of scalars and for the vector field:
\begin{eqnarray} \label{ten}
\Lambda_{ij}=\frac{1}{2}C_{ij}^Is^I;~~~~(A_{\mu})_{ij}=C_{i;j}^IA_{\mu}^I \, ,
\end{eqnarray} where $C_{ij}^I$ and $C_{i;j}^I$ are traceless symmetric and antisymmetric
matrices providing bases (an upper index) for the irreps $[02]$
and $[20]$, respectively. The normalization properties of these
matrices are discussed in Appendix 2. We set out to work with the
Euclidean version of the AdS metric which results in changing the
overall sign of the Lagrangian.

Let us mention the issue of the overall normalization of the gravity action. We
normalize the action of eleven-dimensional supergravity as
$S=\frac{1}{2k_{11}^2}\int \sqrt{g}R+\cdots $, where $k_{11}$ is the
eleven-dimensional Newton constant: $\frac{1}{2k_{11}^2}=\frac{2N^3}{\pi^5}$.
For the $AdS_7\times S^4$ solution with the radii $R_{AdS_7}=1$ and
$R_{S^4}=1/2$ the reduction to seven dimensions yields
$\frac{1}{2k_{7}^2}=\frac{N^3}{3\pi^3}$.

Thus, in the sequel we will work with the action \begin{eqnarray} \label{act}
S(s)= \frac{N^3}{3\pi^3}\int_{AdS_7}\sqrt{g}\, {\cal L}\,  ,\end{eqnarray}
where, after the manipulations described above the Lagrangian acquires the form
\begin{eqnarray} \nonumber {\cal L}&=&\frac{1}{4}(\nabla_{\mu}s^I
\nabla^{\mu}s^I -8s^Is^I)+ 2\, C_{I_1I_2I_3}s^{I_1}s^{I_2}s^{I_3}
\\
\label{Lag} &-&\frac{1}{4}T_{\m\n}\phi^{\m\n}-{\cal L}_2(\phi_{\m\n})+ 4J_{\mu;
\, I} A^{\mu;\, I}
+\frac{1}{2}\,  F_{\m\n}^IF^{\m\n;\, I}  \\
\nonumber &-&\frac{1}{2^4}C_{I_1I_2I_3I_4} \nabla_{\mu}(s^{I_1}s^{I_2})
\nabla^{\mu}(s^{I_3}s^{I_4}) - \, \frac{1}{2}s^{I_1}s^{I_1}s^{I_2}s^{I_2}
+\frac{5}{2}\, C_{I_1I_2I_3I_4}s^{I_1}s^{I_2}s^{I_3}s^{I_4} \, .\end{eqnarray}
Here we introduced the currents \begin{eqnarray} \nonumber
T_{\m\n}&=&\nabla_{\m}s^I\nabla_{\n}s^I-\frac{1}{2}g_{\m\n}
(\nabla_{\rho}s^I \nabla^{\rho}s^I -8s^Is^I) \, ,\\
\nonumber J_{\mu}^I&=&T^{I_1I_2I}s^{I_1}\nabla_{\mu}s^{I_2} \,
\end{eqnarray} obeying an on-shell conservation law. The tensors
$C^{I_1\ldots I_n}$ were already introduced in Section 2.2, and
$T^{I_1I_2I_3}=C_{ij}^{I_1}C_{jk}^{I_2}C_{k;i}^{I_3}$  is antisymmetric in the
indices $I_1,I_2$. In what follows we use a more concise notation, e.g.,
$C_{I_1I_2I_3I_4}\equiv C_{1234}$. In eq. (\ref{Lag})  ${\cal
L}_2(\phi_{\m\n})$ stands for the standard quadratic Lagrangian of the graviton
$\phi_{\m\n}$.

Now it is straightforward, although rather tedious to compute the on-shell
value of the above action subject to the Dirichlet boundary conditions. Like in
the case of gauged $d=5$ supergravity, we have to evaluate the exchange graphs
\footnote{The exchange AdS graphs are reduced to the contact ones by using the
technique developed in ref. \cite{dHFR}.} describing quartic scalar
interactions \cite{AF,AF2}. We omit the details of the computation since they
are similar to those of ref. \cite{AF}. We present only the final result for
the four-point amplitude of the canonically normalized 1/2 BPS operators which
is found by varying the on-shell action with respect to the boundary data for
the scalars: \begin{eqnarray} \label{4pt} \langle {\cal O}^{I_1}(x_1)\cdots
{\cal O}^{I_4}(x_4)\rangle &=& \frac{\d^{12}\d^{34}}{x_{12}^8x_{34}^8}
\\
\nonumber &+&\frac{2^5\cdot 3^3}{\pi^3N^3}\Biggl[
C^{+}_{1234}A^{+}_{1234}+\d^{12}\d^{34}A^0_{1234}+C^{-}_{1234}A^{-}_{1234}
\Biggr]+\mbox{t}+\mbox{u}\, . \end{eqnarray} Here
$C^{\pm}_{1234}=\frac{1}{2}(C_{1234}\pm C_{2134})$. We exhibit explicitly only the
expression in the s-channel, the t-channel is obtained by replacing
$1\leftrightarrow 4$, and the u-channel by $1\leftrightarrow 3$. The first term
in (\ref{4pt}) and its t- and u-counterparts represent the contributions of the
disconnected AdS graphs. The coefficients  $A^{\pm,0}$ are obtained in terms of
the $D$-functions defined in Appendix 2 and read
\begin{eqnarray} \nonumber
&&A^{+}_{1234}=\frac{1}{2x_{34}^4}D_{4422}
+\frac{1}{x_{34}^2}D_{4433}-\frac{7}{2}D_{4444}+4x_{34}^2D_{4455}\, ,   \\
\nonumber
\\
\label{Coeff} &&A^{0}_{1234}=-\frac{1}{6x^4_{34}}D_{4422}
+\frac{1}{18}\left(\frac{x_{13}^2x_{24}^2}{x_{12}^2x_{34}^2}+
\frac{x_{14}^2x_{23}^2}{x_{12}^2x_{34}^2}-\frac{25}{4}\right)\frac{1}{x_{34}^2}D_{4433} \\
\nonumber &&\hspace{1.4cm}
+\frac{1}{8}\left(\frac{x_{13}^2x_{24}^2}{x_{12}^2x_{34}^2}+
\frac{x_{14}^2x_{23}^2}{x_{12}^2x_{34}^2}+\frac{31}{4}\right)D_{4444}
+\frac{1}{2}\left(\frac{x_{13}^2x_{24}^2}{x_{12}^2x_{34}^2}+
\frac{x_{14}^2x_{23}^2}{x_{12}^2x_{34}^2}-1\right)
x^2_{34}D_{4455} \, ,\\
\nonumber
\\
\nonumber &&A^{-}_{1234}=\frac{1}{x_{12}^2x_{34}^2}\Biggl[
(x_{14}^2x_{23}^2-x_{13}^2x_{24}^2)D_{4444}
+\frac{5}{6}(x_{24}^2D_{3434}-x_{14}^2D_{4334}-x_{23}^2D_{3443}+x_{13}^2D_{4343})\Biggr]\\
\nonumber && \hspace{0.6cm}
 + \frac{1}{x_{12}^4x_{34}^4}\Biggl[
\frac{1}{9}(x_{14}^2x_{23}^2-x_{13}^2x_{24}^2)D_{3333}
+\frac{1}{18}(x_{24}^2D_{2323}-x_{14}^2D_{3223}-x_{23}^2D_{2332}+x_{13}^2D_{3232})\Biggr]
\, . \end{eqnarray}

Having found the four-point amplitude, we first check if it obeys the
superconformal Ward identities (\ref{d6con}) derived in Section \ref{sect2}.
Rewriting the amplitude (\ref{4pt}) in the form (\ref{Amp}), we make the
following identification: \begin{eqnarray} \label{ident}
a_1(s,t)&=&1+\frac{2^5\cdot
3^3}{\pi^3N^3}\, x_{12}^8x_{34}^8\, A^0_{1234}\, , \\
\nonumber b_2(s,t)&=&\frac{2^4\cdot 3^3}{\pi^3N^3}\, x_{12}^4x_{14}^4x_{23}^
4x_{34}^4\, (A^+_{1234}+A^+_{3214}+A^-_{1234}+A^-_{3214}) \, . \end{eqnarray}

Now we recall that all $D$-functions appearing in (\ref{Coeff}) can be
expressed as derivatives of $D_{2222}$ with respect to $x_{ij}^2$. On the other
hand, the function $D_{2222}$ itself is given by \begin{eqnarray} D_{2222}=
\frac{\pi^3}{2 x_{12}^2 x_{13}^2x_{24}^2x_{34}^2}(s\pa_s)^2\, \Phi(s,t) \,
,\end{eqnarray} where the function ${\F}(s,t)$, introduced in ref. \cite{UD},
admits the following explicit representation in terms of logarithms and
dilogarithms:
\begin{eqnarray}
\Phi(s,t)=\frac{1}{\l}\left(2(\mbox{Li}_2(-\rho s)+\mbox{Li}_2(-\rho
t))+\ln\frac{t}{s}\ln\frac{1+\rho t}{1+\rho s}+\ln(\rho s)\ln(\rho t
)+\frac{\pi^2}{3}\right)\, .\end{eqnarray} In this way we therefore obtain a
representation for the coefficients $A^{\pm ,0 }$ in terms of certain
differential operators in the variables $s,t$ acting on ${\F}(s,t)$, which is
given in Appendix 3. Such a representation proves useful, since the derivatives
$\pa_s {\F}(s,t)$ and $\pa_t {\F}(s,t)$ are again expressed in a simple manner
via ${\F}(s,t)$. Using the formulae (\ref{A0})-(\ref{Am}) together with
(\ref{ident}), we have verified that the supergravity amplitude we found does
indeed obey the superconformal Ward identities (\ref{d6con}). According to our
general considerations from Section 3, this means that a prepotential of the
type (\ref{Gamplitude}) should exist.

At first sight, the problem of finding the prepotential corresponding to the
supergravity solution (\ref{Coeff}) looks extremely complicated, because one
needs to perform the integral (\ref{Integ}) whose integrand involves
$\Phi(s,t)$. To solve this problem we make the assumption that ${\cal F}$
(\ref{Integ}), written in terms of the variables $s,t$, has the structure
$q_1(s,t)\Phi(s,t)+q_2(s,t)$, where $q_1$ and $q_2$ are two unknown symmetric
functions. Then using the fact that the derivatives of ${\F}$ are again
expressed via ${\F}$ and by trial and error we were able to find these unknown
functions. The final answer is surprisingly simple
 \bea
\label{prepsimple} {\cal F}(s,t)=\frac{\l^3}{2 N^3\, st}
(1-s\pa_s)(1-t\pa_t)(2+s\pa_s+t\pa_t)(1+s\pa_s+t\pa_t)(st\pa_{st})\F(s,t)\,.
\eea We can now  directly verify that substituting (\ref{prepsimple}) in eqs.
(\ref{Gamplitude}) reproduces exactly the coefficients $a_i$ of the four-point
amplitude. To reproduce $b_i$ from the prepotential (\ref{prepsimple}) as well,
we found that a particular value of the integration constant $B$ is required,
namely $B=\frac{1}{N^3}$. Since $\Phi(s,t)$ obeys the crossing symmetry
relation $\Phi\left(s/t,1/t\right)=t\Phi(s,t)$, one can prove that the same
relation holds for the prepotential ${\cal F}(s,t)$, in accord with our
previous considerations.

The form (\ref{prepsimple}) suggests that it can be recast in
terms of the so-called $\bar{D}$-functions that are defined in
Appendix 3. Indeed, it is easy to see that the following formula
holds
\begin{eqnarray} \label{prepfinal} {\cal F}(s,t)=\frac{240}{N^3}\,
\frac{\l^3}{st^2}\, \bar{D}_{7333}(s,t) \, .
\end{eqnarray} This time the crossing symmetry relations for the
prepotential follow from the ones for the corresponding $\bar{D}$-function,
eqs. (\ref{tr1}) and (\ref{tr2}) from Appendix 3. Such an elegant form of the
prepotential suggests that there may exist a much simpler way of extracting it
from the supergravity solution.

Thus we have completely unraveled the structure of the
supergravity solution. It consists of the prepotential
(\ref{prepfinal}) supplemented with the following integration
constants $A$, $B$:
\begin{eqnarray} A=1\, , \qquad B=\frac{1}{N^3} \, . \end{eqnarray}
The comparison of this result with the one provided by the free theory has
already been discussed in Section \ref{sect1.3}.

Finally, we establish a non-trivial relation between the $d=6$ prepotential
(\ref{prepsimple}) and its four-dimensional analogue (\ref{pot4}). The function
$F(s,t)$ from the solution (\ref{pot4}) can be written in a form similar to
(\ref{prepsimple}): \bea F(s,t)=-\frac{4}{N^2}(1+s\pa_s+t\pa_t)(st\pa_{st})\,
\F(s,t) \, . \eea Therefore, comparing with (\ref{prepsimple}) we obtain the
following relation \bea \label{dress} {\cal F}(s,t)={\cal D}_{s,t}F(s,t)\, ,
\eea where  \bea \nonumber {\cal D}_{s,t}= -\frac{1}{8N}\frac{\l^3}{st}
(1-s\pa_s)(1-t\pa_t)(2+s\pa_s+t\pa_t)  \eea  is a symmetric third-order
differential operator. One can easily check that it satisfies the commutation
relation ${\cal D}_{s/t,1/t}\cdot t = t\cdot {\cal D}_{s,t}$ which makes the
crossing symmetry relation for ${\cal F}(s,t)$ obvious. Thus, we conclude that
at large $N$ the dynamical properties of the stress-tensor multiplet in (2,0)
theory are inherited from those of the $d=4$ ${\cal N}=4$ theory.

The last observation suggests another non-trivial test of the original AdS/CFT
duality conjecture for the $d=4$ ${\cal N}=4$ theory. In perturbation theory
the $d=4$ function $F(s,t)$ appears  as a series $
F(s,t)=\frac{1}{N^2}F_1(s,t;g)+{\cal O}\left(\frac{1}{N^4}\right) \,$, where
$g=g_{YM}^2N$ is the t'Hooft coupling. If we assume that $F_1(s,t;g)$
interpolates  smoothly between the free theory ($g=0$) and the theory dual to
the corresponding supergravity ($g=\infty$), then formula (\ref{dress})
provides a smooth deformation connecting the free (2,0) theory and its
supergravity dual. However the OPE of the 1/2 BPS operators in the free (2,0)
theory contains the Konishi-like multiplets while the corresponding
supergravity OPE does not. Therefore, the decoupling of the {\it protected}
Konishi-type multiplets along this particular deformation flow induced by the
${\cal N}=4$ theory should take place only due to the vanishing of their OPE
coefficients. Here we should recall the $d=4$ case, where the Konishi-like
multiplets decouple because their conformal dimensions tends to infinity. Thus,
the known one- and two-loop results for $F_1(s,t;g)$ could be analyzed to see
whether under (\ref{dress}) the protected Konishi-type multiplets emerge or
not. Actually, it would be very interesting to understand how the
supermultiplets arising in the OPE of the $\ell=2$ 1/2 BPS operators of the
${\cal N}=4$ $d=4$ theory are rearranged under (\ref{dress}), as well as  to
clarify the meaning of the operator ${\cal D}_{s,t}$ both on the AdS and the
CFT sides.

\section{Operator Product Expansion}\label{sect4}

In Section \ref{sect1.1} we have already presented the general kinematical
restrictions on the OPE content of two 1/2 BPS operators ${\cal D}(4;000;02)$.
At the level of the four-point amplitude these restrictions are encoded in the
solution (\ref{Gamplitude}) of the superconformal Ward identities. In
principle, one should be able to obtain the conformal partial wave expansion of
the four-point amplitude (\ref{Gamplitude}) with an arbitrary function ${\cal
F}(s,t)$ and to restore all the information about the OPE which was obtained
from solving the kinematical constraints. \footnote{For the $d=4$ ${\cal N}=4$
theory the corresponding conformal partial wave analysis was performed in ref.
\cite{AEPS}.} However, the presence of the second-order differential operator
$\Delta$ in (\ref{Gamplitude}) complicates the analysis considerably and we
have not yet found an easy way to do it. Therefore, in this paper we confine
ourselves to the study of the conformal partial wave expansion of the
particular supergravity amplitude (\ref{4pt}).

%We start by recalling the dimensions of the R symmetry irreps which enter the
%OPE of two stress-tensor multiplets:
%$$\dim[02]=14\,, \quad \dim[20]=10\,, \quad \dim[04]=55\,, \quad \dim[40]=35\,,
%\quad \dim[22]=81 \,.$$

The leading terms in the double OPE arising in the short-distance limit $x_1\to
x_2$, $x_3\to x_4$ can be found as follows. First we project the four-point
amplitude (\ref{4pt}) on the different R symmetry channels (the necessary
projectors are given in Appendix 2). Then we replace the $D$-functions by their
series representation (with powers and logs). The series representation of an
arbitrary $D$-function was worked out in detail ref. \cite{AFP1}. In
particular, the short distance limit under consideration is naturally described
in terms of the variables:
$$
v=\frac{s}{t}\, , \qquad Y= 1 - \frac{1}{t} \, ,
$$
such that $v\to 0$, $Y\to 0$. The leading term
$$
v^{\frac{\tau}{2}}F(Y)
$$
in the conformal partial wave amplitude (CPWA) expansion of the four-point
amplitude corresponds to the contributions of all operators of twist
$\tau=\ell-s$. A logarithmic term of the form $v^{\frac{\tau}{2}}Y^s\ln v$
signals an anomalous dimensions for an operator of twist $\tau$ and spin $s$.
In the sequel we will work out in detail {\it only the leading terms} of the
conformal partial wave expansion for (\ref{4pt}) for the singlet (unprotected)
R symmetry channel and briefly comment on what we have found in the remaining
(protected) channels.

\subsection{Projection on the singlet}

The projection of the connected part of the four-point supergravity amplitude
on the R symmetry singlet channel can be schematically written in the form
\begin{eqnarray} \label{singlet}
\langle {\cal O}\cdots {\cal O}
\rangle_{[00]}=\frac{\d^{12}\d^{34}}{N^3x_{12}^8x_{34}^8}
\Biggl[\frac{12}{175}v^2F_2(Y)+\frac{12}{175}v^3F_3(Y)+v^4\log v \,
G_4(Y)\Biggr], \end{eqnarray}
%\begin{eqnarray}F_2(Y)=
%\frac{12}{175}t_2(Y),~~~~F_3(Y)=\frac{12}{175}t_3(Y)\, .
%\frac{1}{40Y^5}\left(Y(60-60Y+11Y^2)-3(-20+30Y-12Y^2+Y^3)\log(1-Y)\right)\, . \end{eqnarray}
where the functions $F_2(Y)$ and $F_3(Y)$ coincide with the canonically
normalized CPWA of a second-rank tensor with $\ell=6$. In particular, the
corresponding power series expansions start as  follows:
\begin{eqnarray} \nonumber F_2(Y)= \frac{1}{4}Y^2+\frac{1}{2}Y^3 + \cdots\, , \qquad
F_3(Y)= -\frac{1}{6}-\frac{1}{4}Y + \cdots \, . \end{eqnarray} This leading
operator is nothing but the stress tensor. Since the $\log v$ term appears only
at order $v^4$, we conclude that the stress tensor keeps its canonical
dimension. The function $G_4(Y)=-\frac{12}{7}+O(Y)$ which implies that the
first operator receiving anomalous dimension $\ell_a$ of order $N^{-3}$ is a
scalar of approximate dimension $\ell=8$. Taking into account the disconnected
part of the supergravity amplitude, one finds
$$
\ell_a=-\frac{24}{N^3} \, .
$$
This perfectly agrees with the classifications of the UIRs presented in Section
\ref{sect1.1}: The superconformal primary operator of canonical dimension
$\ell=8$ lies beyond the unitarity bound of the continuous series A and is
allowed to acquire an anomalous dimension in a non-trivial interacting theory.
With the help of the techniques developed in refs. \cite{HMMR,DO} the
calculation of the anomalous dimension could be extended to the higher
supermultiplets occurring in the R symmetry singlet channel. It is then of
interest to see how these anomalous dimensions are related to those of the
${\cal N}=4$ theory.

\hspace{0.5cm} It is also instructive to make the comparison with the free
theory of $\eta$ (2,0) tensor multiplets. One tensor multiplet comprises five
scalars, two Weyl fermions and a two-form with a self-dual field strength.
Assuming the free form of the propagator $ \frac{\d^{ij}\d_{ab}}{\pi^3
x_{12}^{4}}$ for the scalars $\phi^i_a$, where $i=1,\ldots, 5$ and $a=1,\ldots,
\eta$, the canonically normalized BPS operator under consideration is of the
form ${\cal O}^I=(2\eta)^{-1/2}\pi^3C_{ij}^I :\phi^i_a\phi^j_a:$. The leading
terms of the corresponding free OPE in the R symmetry singlet channel are
\begin{eqnarray} \nonumber
{\cal O}^{I_1}(x_1){\cal O}^{I_2}(x_2)&=&\d^{I_1I_2}\Biggl[\frac{1}{x_{12}^8}
+\Biggl(\frac{8}{5\eta}\Biggr)^{1/2}[K]-\frac{4\pi^3}{5\eta}
\frac{x_{12}^{\mu}x_{12}^{\nu}}{x^4_{12}}[T_{\mu\nu}^s] +\ldots \Biggr]
\label{OPEfree} \end{eqnarray} where $K=
\pi^3(10\eta)^{-1/2}:\phi^i_a\phi^i_a:$ is a canonically normalized bilinear
operator which we can call a ``Konishi-type scalar" and \begin{eqnarray}
\label{split} T_{\mu\nu}^s=\pa_{\mu}\phi^i_a\pa_{\nu}\phi^i_a-\frac{1}{5}
\pa_{\mu}\pa_{\nu}(\phi^i_a\phi^i_a)-\frac{1}{10}\d_{\mu\nu}(\pa_{\rho}\phi^i_a
\pa_{\rho}\phi^i_a) \end{eqnarray} is the stress-tensor of $5\eta$ free scalars
normalized as $\langle T^sT^s\rangle=\frac{6}{\pi^6}\eta$. The brackets
$[\cdots ]$ denote the contribution of the conformal block of the corresponding
primary operator. It is easy to see, however, that in the free theory the
operator $T^s$ can be written as a sum of three operators $T_{\m\n}$,
$K_{\m\n}$ and $\Sigma_{\m\n}$ which belong to different supermultiplets:
\begin{eqnarray}\label{splitt}
T^s_{\m\n}=\frac{1}{14}T_{\m\n}+\frac{25}{42}K_{\m\n}+\frac{1}{3}\Sigma_{\m\n}
\, . \end{eqnarray} The operators on the right-hand side are orthogonal to each
other, i.e., the mixed  two-point functions vanish. $T_{\m\n}$ is the stress
tensor of $\eta$ copies of the $(2,0)$ theory, $K_{\m\n}$ belongs to the
Konishi-type multiplet and $\Sigma_{\m\n}$ is the leading component of a new
current multiplet.

In fact, the operators $K$ and $\Sigma_{\m\n}$ are the first two
operators from an infinite tower of Konishi-type currents arising
in the singlet channel of the free OPE, all of them having twist
$\tau=4$. However, as we have shown above, the only operator of
$\tau=4$ contributing to the CPWA expansion of the supergravity
four-point amplitude and thus to the OPE, is the stress tensor.
Therefore, all the Konishi-type currents are absent in the
supergravity OPE. Unlike the $d=4$ ${\cal N}=4$ theory, in $d=6$
unitarity puts all of these currents in the isolated series B of
UIRs, so they cannot develop an anomalous dimension in the
interacting theory. We therefore arrive at our conclusion about
the absence of a superconformal theory smoothly interpolating
between the free theory and the one described by the supergravity
dual. Apart from this, the free and the supergravity-dual (2,0)
theories have exactly the same features as their counterparts in
$d=4$. In particular, the same type of splitting (\ref{splitt})
for $T^s$ occurs in $d=4$ \cite{Anselmi,AFP,AFP1}, which merely
reflects the similar structure of the supersymmetry algebras in
$d=4$ and $d=6$.

\hspace{0.5cm} Finally, we comment once more on the relationship of our results
with those obtained in refs. \cite{K,BFT1,BFT2}. If we substitute the
splitting (\ref{splitt}) into the free OPE, then the coefficient in front of
the stress tensor, which equals $C_{{\cal O}{\cal O}T}/C_T$, becomes
$\frac{2\pi^3}{35\eta}$. Here $C_T=\langle TT\rangle$ is the coefficient of the
two-point function of the stress tensor and $C_{{\cal O}{\cal O}T}$ is the
normalization constant of the three-point function of two scalars ${\cal O}$
with the stress tensor. According to ref. \cite{BFT2}, one has
$C_T=\frac{84}{\pi^6}\eta$ and, therefore, we get $C_{{\cal O}{\cal O}T}
=\frac{24}{5\pi^3}$. The same value also follows from the conformal Ward
identity relating the three-point function $\langle {\cal O}{\cal O}T \rangle$
to the two-point function $\langle {\cal O}{\cal O} \rangle$. Hence, the
coefficient in front of the canonically normalized CPWA of the stress tensor in
the CPWA expansion of the four-point amplitude turns out to be $C_{{\cal
O}{\cal O}T}^2/C_T=\frac{48}{175\eta}$. If we want to match it with the
supergravity result (\ref{singlet}), i.e., with the value $\frac{12}{175N^3}$,
we should choose the number of free multiplets to be $\eta=4N^3$. This is of
course a manifestation at the OPE level of the equality between the free and
the supergravity dual integration constants $A,B$.

\subsection{Projection on [02]}

This projection gives \begin{eqnarray} \label{14} \langle {\cal O}\cdots {\cal
O} \rangle_{[02]}= \frac{C^{12}_{{\cal J}_{[02]}}C^{34}_{{\cal
J}_{[02]}}}{N^3x_{12}^8x_{34}^8}
\Biggl[\frac{27}{10}v^2F_2(v)+\frac{27}{10}v^3F_3(Y)+v^4F_4(Y)+v^4\log v \,
G_4(Y)\Biggr], \end{eqnarray} Here and in what follows $C^{12}_{\cal J}$ denote
the orthonormal Clebsh-Gordon coefficient for an irrep ${\cal J}$ appearing in
the tensor product $[02]\times [02]$ (recall (\ref{tp6})). In (\ref{14}) the
functions
\begin{eqnarray} \nonumber F_2(Y)=1+Y+\cdots\, ,
\qquad \nonumber F_3(Y)=\frac{2}{5}+\frac{3}{5}Y+\cdots \,.
\end{eqnarray}
represent the contribution of the canonical CPWA of the $\ell=4$ scalar which
is nothing but the 1/2 BPS primary operator ${\cal O}$. We also find
$F_4(Y)=\frac{97}{175}+O(Y)$, where the first term receives, in particular, a
contribution from a scalar of free field dimension $\ell=8$. On the other hand,
the function $G_4(Y)$ has the form
$$ G_4(Y)=-\frac{27}{7}Y^2+O(Y^3)
\, . $$ Since $G_4$ does not contain a constant term, we conclude that the
scalar of free field dimension 8 transforming in the $[02]$ does not receive
corrections to its free field dimension. According to the classification of
UIRs, this scalar gives rise to a semishort multiplet from the isolated series
B. Recall that in the $d=4$ case the corresponding operator is also a protected
semishort multiplet, however, there unitarity puts it at the bound of the
continuous series A. Its protection can be understood as a consequence of the
three-point function selection rules. The first operator in (\ref{14})
receiving an anomalous dimension is a second-rank tensor of approximate
dimension 10.

\subsection{Projection on [20]}
This projection gives
\begin{eqnarray}
\nonumber \langle {\cal O}\cdots {\cal O} \rangle_{[20]}= \frac{C^{12}_{{\cal
J}_{[20]}}C^{34}_{{\cal J}_{[20]}}}{N^3x_{12}^8x_{34}^8}
\Biggl[\frac{7}{10}v^2F_2(v)+\frac{7}{10}v^3F_3(Y)+v^4\log v \, G_4(Y)\Biggr],
\end{eqnarray} where
$$
F_2(Y)=Y+\frac{3}{2}Y^2+\cdots \, ,  \qquad
F_3(Y)=\frac{3}{7}Y+\frac{6}{7}Y^2+\cdots \,
$$
are precisely the contributions of the CPWA of the $\ell=5$ R symmetry current.
The function $G_4(Y)=-8Y+O(Y^2)$, therefore the first operator in the $[20]$
receiving an anomalous dimension is a vector of approximate dimension 9.

\subsection{Projection on [40]}
This projection gives
\begin{eqnarray}
\nonumber \langle {\cal O}\cdots {\cal O} \rangle_{[40]}= \frac{C^{12}_{{\cal
J}_{[40]}}C^{34}_{{\cal J}_{[40]}}}{N^3x_{12}^8x_{34}^8}
\Biggl[v^4F_4(v)+v^5F_5(Y)+v^5\log v \, G_5(Y)\Biggr]\, . \end{eqnarray}
Therefore, all traceless symmetric rank-$2k$ tensors of twist $8$ transforming
in the $[40]$ are non-renormalized. The explicit form of the function $G_5$
shows that the first operator acquiring an anomalous dimension is a scalar of
approximate dimension $10$.

\subsection{Projection on [04]}
This projection gives
\begin{eqnarray}
\nonumber \langle {\cal O}\cdots {\cal O} \rangle_{[04]}= \frac{C^{12}_{{\cal
J}_{[04]}}C^{34}_{{\cal J}_{[04]}}}{N^3x_{12}^8x_{34}^8}
\Biggl[v^4F_4(v)+v^5F_5(Y)+v^6\log v \, G_6(Y)\Biggr] \, . \end{eqnarray} Such
a structure shows that all rank-$2k$ tensors of twists $8$ and $10$ are
non-renormalized. \footnote{Note that the channel $[04]$ contains {\it two}
towers of protected tensors. The same behavior occurs for the irrep $[040]$ in
the $d=4$ theory \cite{AFP}.} The first operator receiving an anomalous
dimension is a scalar of approximate dimension $12$.

\subsection{Projection on [22]}
This projection gives
\begin{eqnarray}
\nonumber \langle {\cal O}\cdots {\cal O} \rangle_{[22]}= \frac{C^{12}_{{\cal
J}_{[22]}}C^{34}_{{\cal J}_{[22]}}}{N^3x_{12}^8x_{34}^8}
\Biggl[v^4F_4(v)+v^5F_5(Y)+v^5\log v \, G_5(Y)\Biggr] \, . \end{eqnarray} The
function $F_4(v)$ comprises contributions of rank-$(2k+1)$ tensors of $\tau=9$.
Since the term $v^4\log v$ is absent, all these tensors are non-renormalized.
The first operator with anomalous dimension turns out to be a vector of
approximate dimension 11.

This completes our OPE analysis. We see that there are several towers of
traceless symmetric tensors in the irreps $[40]$, $[04]$ and $[22]$ with
vanishing anomalous dimensions. This is again in complete agreement with the
classification of the UIRs, because the corresponding superconformal primary
operators belong to the isolated series D, i.e., they are BPS short. 
Operators with anomalous dimensions in the first five channels of (\ref{tp6})
are supersymmetry descendents of the superconformal primary operators in the R symmetry
singlet. 
As to the
OPEs in the supergravity regime, we conclude that the $d=4$ and $d=6$ theories
have an identical structure though their relation with the free field limit is
of completely different nature.

\section*{Acknowledgements} We are grateful to F. Bastianelli, S. {}Ferrara, P. {}Fr\'e, S. {}Frolov, S. Minwalla,
A. Petkou, K. Skenderis, S. Theisen and A. Tseytlin for useful discussions. G.
A. would like to thank the CERN Theory Division, where part of this work was
done, for the kind hospitality. G. A. was supported by the DFG and by the
European Commission RTN programme HPRN-CT-2000-00131, and in part by RFBI grant
N99-01-00166 and by INTAS-99-1782.

\section{Appendix 1}

For completeness here we recall the general solution of the $d=4$ Ward
identities \cite{EPSS}. The integrability condition for the system
(\ref{d4con}) reads \bea \left[
s\pa_{ss}+t\pa_{tt}+(s+t-1)\pa_{st}+2\pa_s+2\pa_t \right]
\left(\frac{a_1}{s}-\frac{a_3}{t} \right)=0 \, . \eea This equation can be
integrated to give \bea a_1(s,t)&=&\frac{s}{\l}[h(\rho s)-h(\rho
t)]+s F(s,t) \,, \\
a_3(s,t)&=&\frac{t}{\l}[h(\rho t)-h(\rho s)]+t F(s,t) \, .\eea Here $F(s,t)$ is
an {\it a priory} arbitrary symmetric function and $h$ is a function of a
single variable. In particular, the constant (free) solution $a_1=a_3=A$
corresponds to \bea h_0(\rho s)=\frac{A}{2}\left(\rho s+\frac{1}{\rho
s}\right)\, , \qquad F_0(s,t)=\frac{A}{2}\frac{s+t}{st}\, .
 \eea
Integrating the equation for $b_2(s,t)$ one gets \bea b_2(s,t)=B-[h(\rho
t)+h(\rho s)]+(1-s-t)F(s,t) \, . \eea Shifting the solution by the free values,
$h\to h+h_0$, $F\to F+F_0$ leaves $b_2(s,t)$ unchanged. Separating the trivial
solution and regarding the remaining $h$ and $F$ as independent functions, the
crossing symmetry relation for the four-point amplitude takes the form
(\ref{crosym0}) for $F$ together with the following condition on $h$: \bea
\left[h(\rho s)+h(\rho t)\right]_{s\to s/t,t\to 1/t}=\mbox{const} \, . \eea
This completes our discussion of the $d=4$ Ward identities.

\section{Appendix 2}

The matrices $C_{ij}^I$ and $C_{m;l}^I$ introduced in eq. (\ref{ten}) are
subject to the following normalization condition:
\begin{eqnarray} \sum_I
C_{ij}^IC_{kl}^I&=&\frac{1}{2}\d_{ik}\d_{jl}+\frac{1}{2}\d_{il}\d_{jk}-
\frac{1}{5}\d_{ij}\d_{kl} \, , ~~~~ \sum_I
C_{i;j}^IC_{k;l}^I=\frac{1}{2}(\d_{ik}\d_{jl}-\d_{il}\d_{kj}) \, .
\end{eqnarray}

The projectors $P_{1234}^{{\cal J}}$ projecting the four-point amplitude
(\ref{Amp}) on the contributions of different R symmetry irreps $\cal J $ are
constructed by using the technique of ref. \cite{AFP}. We find the following
formulae:
\begin{eqnarray} P_{1234}^{\bf{[00]}}&=&\frac{1}{196}\d_{12}\d_{34}\, , \nonumber\\
P_{1234}^{\bf{[02]}}&=&\frac{10}{189}\left(C_{1234}^{+}-\frac{1}{5}\d_{12}\d_{34}\right)\, ,\nonumber\\
P_{1234}^{\bf{[20]}}&=&-\frac{2}{35}C_{1234}^{-} \, ,\nonumber\\
P_{1234}^{\bf{[40]}}&=&\frac{1}{105}\left(\d_{13}\d_{24}+\d_{14}\d_{23}
+\frac{1}{6}\d_{12}\d_{34}-2C_{1324}-\frac{4}{3}C_{1234}^+\right) \, ,\nonumber\\
P_{1234}^{\bf{[04]}}&=&\frac{1}{330}\left(\d_{13}\d_{24}+\d_{14}\d_{23}
+\frac{8}{63}\d_{12}\d_{34}+4C_{1324}-\frac{16}{9}C_{1234}^+\right) \, ,\nonumber\\
P_{1234}^{\bf{[22]}}&=&\frac{1}{162}\left(\d_{13}\d_{24}-\d_{14}\d_{23}
+\frac{8}{7}C_{1234}^{-}\right) \, .
 \nonumber\end{eqnarray}
In particular, the projectors are normalized to obey the condition
$(P_D)^2={1}/{\nu_{\cal J}}$, where $\nu_{\cal J}$ is the dimension of the
representation $\cal J$: $$\dim[00]=1\,, \quad \dim[02]=14\,, \quad
\dim[20]=10\,, \quad \dim[04]=55\,, \quad \dim[40]=35\,, \quad \dim[22]=81\,.$$

When working out the action of the projection operators on the four-point
amplitude (\ref{Amp}), the following contractions prove helpful:
\begin{eqnarray}
\begin{array}{lll}
C_{1234}C_{1234}=\frac{3199}{50}, & C^{+}_{1234}C_{1234}^{+}=\frac{6671}{200}, & C^{+}_{1234}C_{1324}=\frac{273}{100},\\
                                  &                                            &                         \\
C_{2134}C_{1234}=\frac{273}{100}, & C^{-}_{1234}C_{1234}^{-}=\frac{245}{8},    & C^{-}_{1234}C_{1324}=0 ,  \\
                                  &                                            &                           \\
C_{1122}=\frac{196}{5},           & C_{1212}=\frac{21}{5} \, .                 &                           \\
                                  &                                            &
\end{array}
\end{eqnarray}

\section{Appendix 3}

The $D$-functions related to the space $AdS_7$ can be defined by the formula
\begin{eqnarray} D_{\D_1\D_2\D_3\D_4}(x_1,x_2,x_3,x_4)=\int \frac{d^6 w
dw_0}{w_0^7} \prod_i K_{\D_i}(x_i,w) \, \,  ,\end{eqnarray} where
$K_{\D}(x,w)=\left(\frac{w_0}{w_0^2+(\w-x)^2}\right)^{\D}$ and the integral is
taken over the seven-dimensional space parametrized by $w=(w_0,\w)$, $\w$ being
a six-dimensional vector.

We also define the $\bar{D}$-functions \cite{AFP} for dimension $d$ which are viewed
here as functions of the conformal cross-ratios $s,t$
\bea
\label{barD}
&&D_{\D_1\D_2\D_3\D_4}(x_1,x_2,x_3,x_4)= \\
\nonumber &&\hspace{2cm}= \frac{\pi^{\frac{d}{2}}\,
\bar{D}_{\D_1\D_2\D_3\D_4}(s,t)}
{(x_{12}^2)^{\frac{\D_1+\D_2-\D_3-\D_4}{2}}(x_{13}^2)^{\frac{\D_1+\D_3-\D_2-\D_4}{2}}
(x_{23}^2)^{\frac{\D_2+\D_3+\D_4-\D_1}{2}}  (x_{14}^2)^{\D_4}} \,
. \eea These functions have the following transformation
properties \bea \label{tr1}
\bar{D}_{\D_1\D_2\D_3\D_4}(t,s)&=&\left(\frac{t}{s}\right)^{\frac{1}{2}(\D_1-\D_2-\D_3-\D_4)}
\bar{D}_{\D_1\D_4\D_3\D_2}(s,t) \, , \\
\label{tr2}
\bar{D}_{\D_1\D_2\D_3\D_4}\left(\frac{s}{t},\frac{1}{t}\right)&=&t^{\frac{1}{2}(\D_1-\D_2-\D_3-\D_4)}
\bar{D}_{\D_1\D_2\D_4\D_3}(s,t) \, , \eea which can be easily proven by using,
e.g., the Feynman parameter representation.

We find the following representations for the coefficients $A^{\pm,0}$ of the
four-point amplitude in terms of differential operators acting on $\Phi(s,t)$:
\begin{eqnarray} \label{A0}
A^{0}_{1234}&=&\frac{\pi^3}{(x_{12}^2x_{34}^2)^3x_{13}^2x_{24}^2}\Biggl[
-\frac{1}{2^3\cdot 3^3}(2-s\pa_s)(1-s\pa_s)(s\pa_s)^2\\
\nonumber &+&\frac{1}{2^5\cdot 3^3}\left(\frac{1}{s}+
\frac{t}{s}-\frac{25}{4}\right) (2-s\pa_s)(1-s\pa_s)^2(s\pa_s)^2\\
\nonumber &+&\frac{1}{2^5\cdot 3^3}\left(\frac{1}{s}+ \nonumber
\frac{t}{s}+\frac{31}{4}\right)(2-s\pa_s)^2(1-s\pa_s)^2(s\pa_s)^2\\
\nonumber &+&\frac{5}{2^7\cdot 3^3}\left(\frac{1}{s}+
\frac{t}{s}-1\right)(3-s\pa_s)(2-s\pa_s)^2(1-s\pa_s)^2(s\pa_s)^2\Biggr]\Phi(s,t)
\, , \\
\label{Ap}
A^{+}_{1234}&=&\frac{\pi^3}{(x_{12}^2x_{34}^2)^3x_{13}^2x_{24}^2}\Biggl[\frac{1}{2^3\cdot
3^2}(2-s\pa_s)(1-s\pa_s)(s\pa_s)^2\\
\nonumber &+&\frac{1}{2^4\cdot 3}(2-s\pa_s)(1-s\pa_s)^2(s\pa_s)^2
-\frac{7}{2^3\cdot 3^3}(2-s\pa_s)^2(1-s\pa_s)^2(s\pa_s)^2 \\
\nonumber &+&\frac{5}{2^4\cdot 3^3}(3-s\pa_s)(2-s\pa_s)^2(1-s\pa_s)^2(s\pa_s)^2
\Biggr]\Phi(s,t)\, ,
\\
\label{Am}
A^{-}_{1234}&=&\frac{\pi^3}{(x_{12}^2x_{34}^2)^3x_{13}^2x_{24}^2}\Biggl[
\frac{1}{2^2\cdot 3^3
}\left(\frac{t}{s}-\frac{1}{s}\right)(2-s\pa_s)^2(1-s\pa_s)^2(s\pa_s)^2
\\
\nonumber
&+&\frac{5}{2^4\cdot 3^2}(1+s\pa_s+2t\pa_t)(1-s\pa_s)^2(s\pa_s)^2\\
\nonumber &+& \frac{1}{2^4\cdot 3^2
}\left(\frac{t}{s}-\frac{1}{s}\right)(1-s\pa_s)^2(s\pa_s)^2 +\frac{1}{2^3\cdot
3^2}(1+s\pa_s+2t\pa_t)(s\pa_s)^2\Biggr]\Phi(s,t) \, . \end{eqnarray} Further
simplification is achieved by successive use of the identities \cite{EPSS}
\begin{eqnarray}
\nonumber \pa_s \Phi(s,t)&=&\frac{1}{\lambda^2}
\left(\Phi(s,t)(1-s+t)+2\ln s -\frac{s+t-1}{s}\ln t\right) \, ,\\
\nonumber \pa_t \Phi(s,t)&=&\frac{1}{\lambda^2} \left(\Phi(s,t)(1-t+s)+2\ln t
-\frac{s+t-1}{t}\ln s\right) \, .
\end{eqnarray}
\newpage

\renewcommand{\baselinestretch}{0.6}

\end{document}